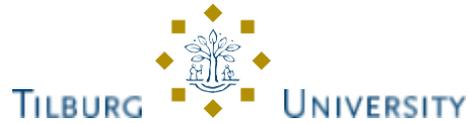

# Modeling the Bioelectric Signature Transition from Healthy Glial Cells to Glioblastoma

## A Computational Approach Using ASAL-like Model to Simulate Bioelectrical State Changes

Wiktoria Agata Pawlak

STUDENT NUMBER
2106820

COMMITTEE
Dr. Travis J. Wiltshire – Supervisor
Ms. Natalie Ranzhi Wei – Second Reader
Dr. Hananel Hazan – External Supervisor Tufts University
Dr. Wesley Clawson - External Supervisor Tufts University

LOCATION
Tilburg University
School of Humanities and Digital Sciences
Department of Cognitive Science & Artificial Intelligence
Tilburg, The Netherlands

DATE
May 15th, 2025

WORD COUNT
8399

ACKNOWLEDGMENTS
I would like to thank my supervisors, Dr. Travis J. Wiltshire (Tilburg
University), Dr. Hananel Hazan, and Dr. Wesley Clawson (Tufts University),
for their valuable feedback throughout this thesis progress. I am also grateful
to Prof. Michael Levin (Tufts University) for the inspiration his research on
bioelectricity provided — whose work, and hours of podcasts I have listened
to, inspired me to pursue this direction — and to Dr. Teodor Emanuel
Buchner (Warsaw University of Technology) for the many thought-provoking
conversations, encouragement, and perspectives that fueled my excitement
for future work and laid the foundation for further research.

To my family — for endless prayers.
And to God — none of this would have come together without Your Word.

"Hear, O Israel: The LORD our God, the LORD is one."
"Have I not commanded you? Be strong and courageous. Do not be
frightened, and do not be dismayed, for the LORD your God is with you
wherever you go."
— Joshua 1:9

# Modeling the Bioelectric Signature Transition from Healthy Glial Cells to Glioblastoma

A Computational Approach Using ASAL-like Model to Simulate Bioelectrical State Changes

Wiktoria Agata Pawlak

### Abstract

This thesis investigates the bioelectric transition from healthy glial cells to glioblastoma, a lethal brain tumor, focusing on the hypothesis that mitochondrial dysfunction triggers a cascade of depolarization indicating potential tumorigenesis in-vivo studies. Unlike prior studies emphasizing genetic or metabolic drivers, we developed a goal-directed framework inspired by the Automated Search for Artificial Life (ASAL)—originally designed for open-ended exploration of life-like properties—to simulate this transition computationally. Our approach integrates a genetic algorithm, a diffusion model, and an agent-based cellular simulation to evolve cellular "genomes" that dictate bioelectric and metabolic traits Simulations were structured into three experimental stages. In Stage A, we varied mitochondrial efficiency to examine its direct effects on membrane potential. In Stage B, we applied quality-diversity techniques—specifically Multi-dimensional Archive of Phenotypic Elites (MAP-Elites) —to explore the resilience of bioelectric states across a diverse parameter space under fixed mitochondrial function. In Stage C, we conducted a full evolutionary search including mitochondrial parameters to uncover tumor-like attractors. Across all stages, we targeted glioblastoma-relevant properties such as depolarized membrane potentials. Results demonstrate that mitochondrial perturbations initiate bioelectric shifts, with membrane potentials depolarizing from -80 mV to a mean of –35 mV (locally up to –30 mV), alongside the emergence of tumor-like spatial patterns driven by ion dynamics and gap junction coupling. These findings support the hypothesis and highlight bioelectric properties as potential therapeutic targets. While traditional ASAL explores broad, unsupervised dynamics, our goal-directed adaptation narrows the focus to glioblastoma-specific behaviors, enhancing its relevance to cancer research. Limitations include short simulation timescales and the exclusion of immune and angiogenic dynamics, which are important for in vivo GBM progression.

DATA SOURCE, ETHICS, CODE AND TECHNOLOGY STATEMENT

## 1.1 *Source/Code/Ethics/Technology Statement*

*Data Source:* This thesis project did not involve the collection of data from human participants or animals. All data used (e.g., model parameters, biological constants, physiological ranges) were obtained from publicly available scientific literature and databases, as detailed in the Appendix. Additionally, the project uses simulated data generated by a custom-developed genetic algorithm inspired by ASAL (Automated Search for Artificial Life), which was adapted specifically for biological data modeling and cancer discovery (Kumar et al., 2024). Data ownership remains with the original authors and publishers; no additional consent was required, as all data are fully anonymized and publicly accessible.

*Figures:* All figures presented in this thesis were generated entirely by the author using Python libraries such as Matplotlib and Seaborn or manually created with tools like Mermaid.js or Canva.com tools. No externally sourced figures requiring consent were included.

*Code:* The computational modeling code developed in this thesis is original work created by the author. It is a custom implementation of a genetic algorithm inspired by ASAL but specifically redesigned to accommodate biological constraints and simulate transitions relevant to glioblastoma research (Kumar et al., 2024). No code from other studies was directly copied. All dependencies and frameworks are documented in the accompanying requirements.txt. Core libraries include: Python 3.10, NumPy 1.26.4, SciPy 1.11.3, Pandas 2.1.1, Matplotlib 3.8.4, Seaborn 0.13.0, JupyterLab 4.0.9, Brian2 2.6.3, scikit-learn 1.3.2, Numba 0.58.1, Torch 2.1.2, and tqdm 4.66.2. The full source code is available in Google Drive as a zip file at: https://drive.google.com/drive/folders/1sMqmJ0n2773NBJ4JJHxdhIwITxffli?usp=sharing Zip contains README.md explaining how to reproduce the code.

*Technology*: Grammarly.com (online web editor, latest version as of April 2025) was used for language refinement, sentence restructuring, and paraphrasing support. Cursor.io was used for code debugging, and Git was used for version control. The thesis was written and formatted using Google Docs. Scribbr.com was used for reference management. No additional paraphrasing, grammar-checking, or typesetting tools were employed.

## 1  Introduction

Glioblastoma multiforme (GBM) is the most aggressive form of brain cancer, known for its rapid proliferation, invasiveness, and resistance to standard treatments such as surgery, chemotherapy, and radiation. Despite major efforts across oncology and neuroscience, GBM patients face a median survival of only 15 months (Kanderi et al., 2024; Sabouri et al., 2024). Traditionally, GBM research has focused on genetic mutations and metabolic shifts, such as the Warburg effect, where cancer cells favor glycolysis even in the presence of oxygen (Sipos et al., 2025; Kanderi et al., 2024). However, a growing body of evidence highlights the role of bioelectric properties of cells—such as membrane potential ($V_m$), ion gradients, and gap junction coupling—as important regulators of cell behavior, and potentially of tumorigenesis itself (Chernet, 2014; Zhang & Levin, 2025).

In healthy glial cells (precursors to GBM), resting membrane potential is tightly regulated around –80 mV through active transport and selective ion channel activity (Molenaar, 2011). In contrast, GBM cells often exhibit a depolarized $V_m$, typically ranging from –40 mV to –20 mV, which contributes to increased proliferation, loss of contact inhibition, and altered signaling (Molenaar, 2011). Crucially, bioelectric changes are not just downstream symptoms of transformation—they may act as causal triggers. Emerging theories posit the existence of a bioelectric prepattern (Levin, 2014), in which early deviations in $V_m$ and ionic homeostasis predispose cells toward tumor-like phenotypes before any genetic mutation is fixed (Molenaar, 2011; Venkatesh et al., 2019).

One candidate mechanism behind this shift is mitochondrial dysfunction, which is a well-documented feature of cancer cells, often linked to metabolic reprogramming (Luo et al., 2020). Yet, its role in disrupting bioelectric properties remains underexplored. Mitochondria are central to cellular energetics, generating ATP used to power ion pumps (e.g., Na$^+$/K$^+$-ATPase) that maintain V☐. Reduced mitochondrial efficiency can lead to ionic imbalance, membrane depolarization, and impaired signaling (Guntuku et al., 2016; Srinivasan et al., 2017). Yet, this connection remains poorly understood in computational models of cancer (Carvalho, 2021).

**Research Question:**
To what extent can mitochondrial dysfunction causally induce bioelectric instability and glioblastoma-like phenotypes?

**Sub-questions:**

- **RQ1:** Can reduced mitochondrial efficiency alone cause depolarization and bioelectric instability?

- **RQ2:** Which bioelectric configurations are most resilient or vulnerable to perturbations?

- **RQ3:** Can tumor-like attractor states spontaneously emerge under evolutionary pressure?

To answer these questions, we developed a computational simulation that integrates bioelectric modeling with evolutionary search, structured into three experimental stages. Our system builds on principles from the Automated Search for Artificial Life (ASAL) framework, which was originally designed for open-ended simulation of emergent life-like behaviors (Kumar et al., 2024). In contrast to ASAL's unsupervised, exploratory orientation, we build a goal-directed architecture to evolve glial tissue toward glioblastoma-relevant phenotypes as ASAL's open-ended nature can be too broad for hypothesis-driven research. Our version of the simulation uses a modular design consisting of four core components typical of ASAL architecture:

- **A genetic algorithm (GA)**, which evolves cellular "genomes"—parameter sets encoding ion channel conductances, mitochondrial efficiency (M$_{eff}$), ATP handling and other physiological traits relevant to bioelectric function.

- A **MAP-Elites quality-diversity archive** (used in Stage B), which organizes genomes into a grid of niches based on phenotypic traits (e.g., V☐ stability, entropy) and keeps a diverse set of high-performing solutions rather than a single optimum (Mouret & Clune, 2015).

- **An agent-based bioelectric simulator**, which models the spatiotemporal dynamics of membrane voltage (V☐), ion concentrations (K$^+$, Na$^+$, Cl$^-$, Ca$^{2+}$, H$^+$), intracellular ATP, pH shifts, and gap junction-mediated electrical coupling across a 2D grid of glial cells.

- **A generative diffusion model**, trained on CLIP (Contrastive Language-Image Pre-training) embeddings of high-performing membrane voltage maps (from Stage B), which evaluates genome candidates and steers the search for new ones based on bioelectric similarity to enrich the search space (Radford et al., 2021).

- **A multi-objective fitness function**, which scores tissue simulations based on membrane depolarization, spatial heterogeneity, metabolic-efficiency preservation, electrical homeostasis, and an auxiliary CLIP-based morphology similarity term.

**MAP-Elites** is a quality-diversity algorithm that explores a search space by evolving and preserving the best-performing solutions across a grid of behaviorally distinct niches. **CLIP** (Contrastive Language–Image Pre-training) is a neural network model that learns visual concepts from natural language by training on image–text pairs, enabling it to embed and compare images and text in a shared semantic space. This modular architecture allows us for iterative optimization of bioelectric traits under biologically motivated constraints (Figure 1). To test our hypothesis, we structured the simulation into three distinct experimental stages:

- **Stage A (Causal Induction):** We performed a parameter sweep of mitochondrial efficiency ($M_\mathscr{E}$) with all the other parameters frozen to assess its direct influence on membrane potential. This tests whether reduced mitochondrial efficiency and energy supply alone can produce depolarized states resembling GBM.

- **Stage B (Resilience Mapping):** Here, mitochondrial efficiency was held constant at $M_\mathscr{E} = 1.0$ (healthy baseline). We applied the MAP-Elites algorithm to explore the diversity of stable and unstable bioelectric configurations. To replicate ASAL principles, the CLIP-based diffusion model was used to propose new genome candidates based on embedding similarity to prior high-performing states, enriching the search space with structured variability. This integration acts as a semantically guided mutation operator, guiding the generation process toward phenotypically similar — but genetically novel — candidates, thus complementing the evolutionary search.

- **Stage C (Emergent Tumor States):** We enabled full genome evolution, including mitochondrial traits, to discover attractor states that naturally evolve toward depolarization, high proliferation, and bioelectric instability. This stage simulates the spontaneous emergence of tumor-like behavior under multi-objective selection.

Each stage addresses a specific dimension of the hypothesis: causal induction (Stage A), resilience (Stage B), and emergence (Stage C). Across all stages, the agent-based tissue model remained constant, incorporating experimentally grounded values from the literature for healthy astrocytes and GBM cells (Appendix A, Appendix B). All simulations began from parameterized

GBM-adjacent initial conditions (Appendix B), as we found that fully healthy parameterizations were strongly resistant to depolarization—an effect likely due to their stable resting-state dynamics, which could not be overcome in the short (<24h) simulation windows dictated by GPU limitations.

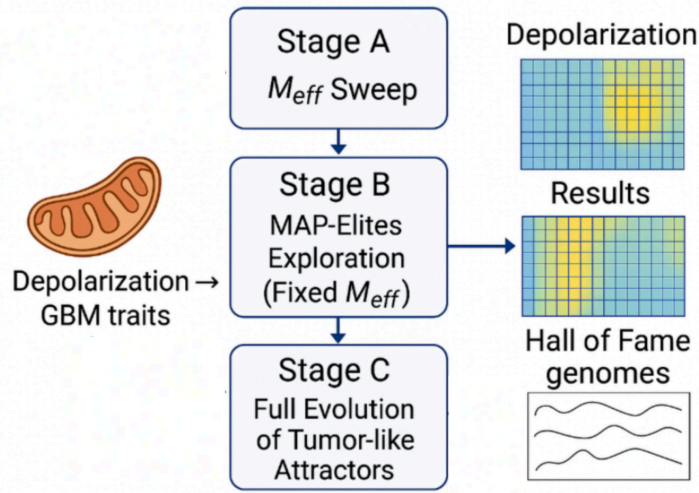

Figure 1. Overview of the three-stage simulation workflow used to model the bioelectric transition in glioblastoma.

The model incorporates essential physiological mechanisms: ATP-driven ion transport, voltage-gated channel kinetics, calcium buffering, proton extrusion, and simplified mesoscopic field coupling. Cell division occurs when metabolic and bioelectric thresholds are met. Temporal oscillations such as circadian and cell-cycle rhythms were also included, although long-term dynamics such as angiogenesis and immune infiltration were not modeled due to computational constraints.

Results from Stage A confirmed that decreasing mitochondrial efficiency induces a shift from –80 mV toward –20 mV in many tissue configurations, consistent with GBM electrophysiology. Stage B showed a structured resilience landscape: some configurations retained polarization despite perturbations, while others collapsed readily into unstable regimes - while still being slightly polarized- but with more difficulty since mitochondrial efficiency was clipped to maximum value. In Stage C, we observed the evolution of tumor-like attractors: clusters of cells with low V☐, disrupted ionic balance, elevated simulated proliferation, and emergent spatial heterogeneity.

This thesis presents a biologically grounded, simulation-based framework to model bioelectric transitions relevant to glioblastoma while still having certain limitations typical of early-stage modeling, such as constrained timescales (<24h for effective mitosis), limited population sizes, limited optimization parameters, limited hyperparameter tuning, and a lack of long-term feedback mechanisms like angiogenesis or immune infiltration. By integrating mitochondrial metabolism, ion transport, and emergent multicellular dynamics with a CLIP-conditioned generative model, we provide mechanistic insight into how bioelectric state instability can emerge from metabolic failure—even in the absence of genetic mutations. This work contributes to the growing field of computational bioelectricity and lays the foundation for future in vivo validation, and eventually, for bioelectrical diagnostics and therapeutic interventions (e.g., electroceuticals or reprogramming protocols), pending greater GPU availability and model scalability.

## 2   Related Work

This section synthesizes relevant research that informed the development of our modular simulation pipeline, organized into five interconnected areas: (1) bioelectricity in cancer, (2) mitochondrial dysfunction and bioelectric instability in glioblastoma (GBM) (3) computational models of tumorigenesis, (4) our adaptations of the Automated Search for Artificial Life (ASAL), and (5) quality-diversity algorithms and generative models in biological simulation. These areas frame the research gap addressed by this study: modeling the bioelectric transition from healthy glial cells to GBM as a consequence of mitochondrial dysfunction, validated by simulation results showing depolarization and ion dysregulation.

### 2.1 Bioelectricity in Cancer

Bioelectricity—defined by membrane potential (V□) and ion flux—regulates key biological processes including embryonic development, tissue patterning, and regeneration. For example, during embryogenesis, spatial gradients in V□ guide cell differentiation and morphogenesis. In regenerative medicine, bioelectric cues can induce limb regeneration in model organisms by influencing stem cell behavior (Levin, 2014). In cancer biology, V□ is equally relevant. It plays a central role in controlling proliferation, migration, and gene expression in glial cells. Healthy glial cells typically maintain a hyperpolarized V□ around −80 mV, sustained by high potassium permeability and $Na^+/K^+$-ATPase activity. In contrast, glioblastoma (GBM) cells exhibit depolarized V□ values between −40 mV and −20 mV, which are associated with increased proliferation, migration, dedifferentiation, and activation of oncogenic pathways (Yang & Brackenbury, 2013; Turner & Sontheimer, 2014).

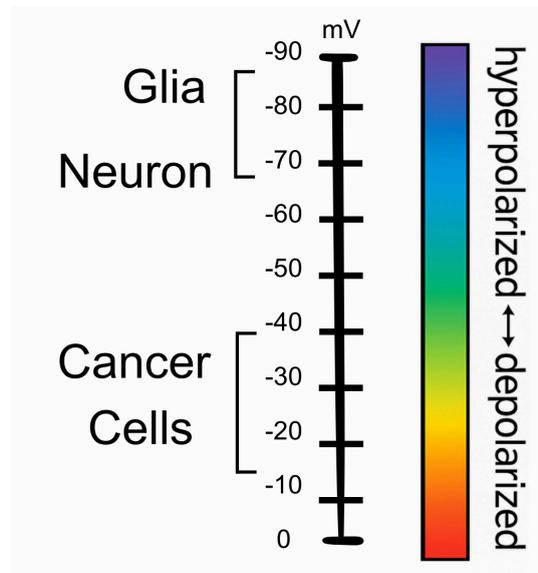

Figure 2. Membrane Potential Ranges in Healthy and Tumor Cells
Resting membrane potentials of neurons, glia, and cancer cells along the
bioelectric polarity spectrum. Glioblastoma and other tumor cells are
characterized by depolarized varying resting potentials, while healthy neurons
and glia maintain more hyperpolarized states.

Increasing evidence suggests shift in $V_m$ is not just a secondary effect of
transformation but may act as a primary driver of tumorigenesis. Early bioelectric
changes can precede genetic mutations, forming what has been termed a
bioelectric prepattern—a physiological state that predisposes cells to malignancy
(Chernet & Levin, 2013; Levin, 2014). Depolarization, for example, activates
voltage-gated ion channels, increases intracellular $Ca^{2+}$ levels, and upregulates
cell cycle genes. GBM cells frequently show an "oncochannel" phenotype,
marked by overexpression of ion channels such as TRPM8 and BK (Turner &
Sontheimer, 2014; Wondergem & Bartley, 2009). Despite these insights, the
upstream triggers of such bioelectric instability remain poorly understood. One
plausible origin is mitochondrial dysfunction, which disrupts ATP production and
impairs ion pump activity, leading to a collapse in $V_m$ stability. Nevertheless,
most computational models of GBM focus heavily on genetic networks, often
overlooking the electrophysiological layer (Fulda et al., 2010; Guntuku et al.,
2016).

In this work, we hypothesize that mitochondrial dysfunction, which impairs ATP
production and disrupts ion pump activity, may initiate bioelectric instability, as
supported by studies linking bioenergetic collapse to cancer progression
(Seyfried, 2015; Seyfried et al., 2013). Parameters such as ion channel
conductances (e.g., $G_{m,leak}$) and ionic concentrations (e.g., $K^+$, $Na^+$, $Ca^{2+}$), which
are central to maintaining $V_m$, may serve as early indicators or contributors to
malignancy when perturbed (Seyfried et al., 2013).

Our computational model simulates the dynamics of glial tissue using these parameters as inputs. While it does not replicate full in vivo tumorigenesis, it enables us to test whether $V_\square$ shifts alone can produce malignant-adjacent electrophysiological signatures. In this way, we operationalize the concept of a bioelectric prepattern and investigate how electrophysiological features—driven by mitochondrial inefficiency and ion transport failure—might initiate or stabilize tumor-like behaviors.

## 2.2 Mitochondrial Dysfunction

Mitochondrial dysfunction is frequently observed in glioblastoma (GBM) and has an important role in both tumor progression and the disruption of bioelectric stability. Mitochondria supply the ATP required to power ion pumps such as the $Na^+/K^+$-ATPase, which maintain membrane potential ($V_\square$) and ionic gradients important for cellular homeostasis (Wallace, 2012). In GBM, impairments in mitochondrial function reduce ATP availability, destabilize ion gradients, and contribute to the Warburg effect—where cells rely on glycolysis rather than oxidative phosphorylation even in the presence of oxygen, resulting in extracellular acidification (Seyfried, 2015; Seyfried et al., 2013). This metabolic compensation enables GBM cells to sustain growth despite defective respiration.

Seyfried and colleagues have extensively argued that cancer, including GBM, is primarily a metabolic disease driven by mitochondrial defects, with genetic mutations as secondary consequences (Seyfried & Shelton, 2010; Seyfried et al., 2013). This metabolic shift is accompanied by increased production of reactive oxygen species (ROS), which further damage mitochondrial membranes, mtDNA, and respiratory enzymes, reinforcing mitochondrial dysfunction and activating oncogenic signaling leading to cascade of metabolic reprogramming (Lue et al., 2017; Sainero-Alcolado et al., 2022; Seyfried et al., 2013). These dysfunctions also alter lipid and amino acid metabolism, which is associated with GBM's aggressive growth and therapeutic resistance.

Mitochondrial dysfunction in GBM is additionally linked to calcium dysregulation. Research by Bikfalvi et al. (2022) shows that mitochondrial calcium overload contributes to tumor progression and resistance to apoptosis, further connecting mitochondrial damage to bioelectric instability. These disruptions can impair calcium-dependent signaling and ion pump function, amplifying depolarization (Giorgi et al., 2018). Studies have shown that such bioenergetic disruptions can precede genetic alterations in cancer development, suggesting a primary role in tumorigenesis (Seyfried & Shelton, 2010; Wallace, 2012). This perspective aligns with bioelectric research, which proposes that early disruptions in $V_\square$ stability may create a bioelectric prepattern, a physiological configuration that predisposes glial cells to malignancy (Levin, 2014). Blackiston et al. (2009) demonstrated that mitochondrial activity influences bioelectric signaling in developmental systems, supporting this mechanistic link between metabolism and electrophysiological state.

We focused on mitochondrial dysfunction as a hypothesized trigger because it may induce early bioelectric changes—such as V▢ depolarization—prior to genetic alterations. Despite these findings, most computational models of GBM prioritize genetic and signaling pathways while overlooking mitochondrial bioenergetics and electrophysiological dynamics.

*2.3 Computational Models and Adaptation of ASAL*

Computational models of cancer have traditionally focused on genetic and evolutionary dynamics. Agent-based models, such as those by Anderson et al. (2006), simulate tumor morphology by emphasizing genetic heterogeneity and clonal evolution, often using spatially explicit frameworks to capture microenvironmental pressures. Similarly, cellular automata frameworks model tumor progression through rule-based state transitions (Basanta & Deutsch, 2008). PhysiCell, a widely used agent-based platform, simulates biomechanical interactions and cell-cell dynamics in 3D multicellular systems but often overlooks bioelectric and metabolic factors important for tumorigenesis (Ghaffarizadeh et al., 2018).

To address these limitations, bioelectric modeling platforms like the BioElectric Tissue Simulation Engine (BETSE) simulate membrane voltage, ion transport, and gap junction coupling in tissue-like environments (Pietak & Levin, 2016). BETSE has been used to study bioelectric signaling in development and cancer, showing how ion flows influence cell behavior and tissue patterning (Manicka & Levin, 2019; Pietak & Levin, 2016). However, BETSE focuses on steady-state simulations and lacks support for evolutionary dynamics or the emergence of novel, unstable configurations. These features are important when combining scattered across fields research findings and understudied tumor behaviors (Stransky & De Souza, 2013). Seeking a more innovative and adaptable approach, we turned to artificial life (ALife) models for inspiration.

ALife models, such as the Automated Search for Artificial Life (ASAL), are designed for open-ended exploration of emergent behaviors like self-replication and pattern formation across diverse substrates (Kumar et al., 2024; Lehman & Stanley, 2010). While ASAL was not developed for biomedical contexts, its capacity to evolve life-like dynamics provided a conceptual foundation for our own modeling framework tailored to cancer research. Rather than using ASAL directly, we built a custom simulator focused on hypothesis-driven exploration of tumorigenesis capable of modeling multicellular bioelectric dynamics, including field propagation and gap junction coupling—features often oversimplified in traditional volume conductor models (Funk & Scholkmann, 2022; Mesin, 2013). Our adaptation retains ASAL's evolutionary search logic but grounds it in biophysics, allowing for targeted exploration of GBM-adjacent states where bioelectric signaling is an important factor in tumor progression (Levin, 2014).

Unlike BETSE, our approach simulates temporally dynamic processes and supports evolutionary optimization through quality-diversity algorithms such as MAP-Elites (Mouret & Clune, 2015). Moreover a genetic algorithm (GA) is ideal for this study because it efficiently explores a vast parameter space to uncover emergent bioelectric and metabolic dynamics in GBM, unlike static models like BETSE or genetic-focused frameworks like PhysiCell. This allows us to uncover rare and potentially pathogenic bioelectric configurations with greater flexibility for hypothesis testing and custom adaptation. This way by bridging bioelectric, metabolic, and genetic layers, our approach attempts to address limitations in both gene-centric and mechanically focused models and cellular automata (Zhang & Levin, 2025) making it a more flexible tool for discovery.

## 3 Method

This chapter outlines the methodology for simulating the bioelectric transition of healthy glial cells into glioblastoma (GBM)-like states, driven by mitochondrial dysfunction. We developed a computational framework inspired by evolutionary algorithms, adapting concepts from the Automated Search for Artificial Life (ASAL) (Kumar et al., 2024) for cancer research. We divided it into three experimental stages. Stage A ran on a 64×64 grid, while Stages B and C used a genetic algorithm (GA) evolving tissue simulations on a 32×32 grid due to computational constraints. It optimized bioelectric and metabolic parameters to replicate GBM electrophysiological signatures, such as depolarized membrane potential ($V\square$), ion dysregulation, and tumor-like clustering. The following sections detail the simulation framework, parameter selection, evolutionary process, fitness function, validation, and experimental design.

### 3.1 Simulation Framework

All three experiments (Stages A–C) use the same agent-based simulation platform. We implemented a 2D agent-based model using a NumPy/Numba-based solver (TissueNP). In Stage A, simulations were run on a 64×64 grid for 60,000 steps (0.05 ms timestep), as the absence of evolutionary computation allowed for higher spatial resolution and shorter runtime. For Stages B and C, we used a 32×32 grid of 1,024 glial cells and ran simulations for 40,000 steps (0.05 ms timestep, totaling 2,000 ms) to balance spatial resolution with computational tractability for evolutionary runs. Each cell is an electrophysiological unit with bioelectric properties (e.g., $V\square$, ion concentrations: $K^+$, $Na^+$, $Ca^{2+}$, $Cl^-$, $H^+$) and metabolic properties (e.g., ATP, mitochondrial efficiency Meff), interacting via gap junctions (conductance G_GAP). The edge bath mode applies Dirichlet boundary conditions to perimeter cells, clamping extracellular $K^+$ and $Na^+$ concentrations, while inner domain bioelectric dynamics evolve freely to model physiological ion buffering by vasculature or CSF at tissue boundaries. Extracellular ion diffusion uses a five-point Laplacian with periodic boundary conditions to ensure mass conservation and avoid artificial edge effects in diffusion dynamics. Figure 2 illustrates the overall algorithmic architecture of the evolutionary simulation. The simulation framework integrates a coupled system of ion channels, pumps (e.g., Na$^+$/K$^+$-ATPase), gap junction coupling, and metabolic modules that update intracellular ATP, ROS, and mitochondrial membrane potential ($\Psi\Box$) based on Meff and ion flux dynamics to model the interdependence between metabolism and electrophysiology.

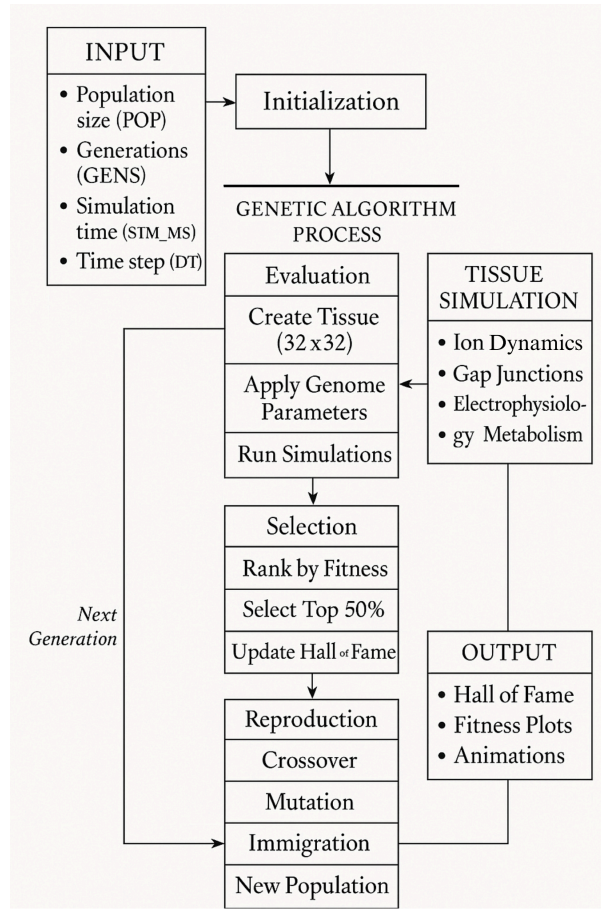

Figure 3. Algorithmic Architecture of the Evolutionary Simulation for stages B and C.

Key components include (*for details about variables please refer to Appendix A*):

- **Bioelectric Dynamics:** Governed by ion channel conductances (e.g., G_K_LEAK, G_NA_VG, G_GAP) and pumps (e.g., Na$^+$/K$^+$-ATPase PUMP_MAX), computed using reversal potentials (E_K, E_Na) and currents (I_K_leak, I_Na_vg). Vm updates use a capacitance-based equation, clamped to ±5 mV per step, with ±0.02 mV Gaussian jitter for stability. Electric field coupling synchronizes Vm across neighboring cells, modeling bioelectric interactions. Implemented in the model to capture core electrophysiological behavior driving GBM-like depolarization using minimal, interpretable parameters.

- **Metabolic Dynamics:** Mitochondrial efficiency (Meff) drives ATP production (oxidative: ATP_OX_RATE, glycolytic: ATP_GLY_RATE). Mitochondrial membrane potential (Psi_m) adjusts based on Meff, influencing ATP synthesis, while reactive oxygen species (ROS) accumulate from mitochondrial inefficiency and calcium overload, decaying over time. ROS exponentially reduces gap junction coupling.

- **Regulation:** Osmotic water flux adjusts cell volume based on intracellular osmolarity, while proton and CO2/HCO3⁻ buffering maintain pH homeostasis via hydration/dehydration reactions and proton exchange. Implemented in the model to maintain homeostasis by accounting for osmotic balance and pH buffering in tumor environments.

- **Agent Behavior:** Cells divide into adjacent empty spaces if ATP exceeds 2.5 mM (ATP_DIV_M), $V_\square$ exceeds -40 mV (VM_DIV_MV), and the cell-cycle clock ($\theta$_cc) exceeds 0.9, halving resources with daughter cells. Depolarized cells ($V_\square > $ -40 mV) migrate toward regions with higher extracellular H⁺ (lower pH), simulating GBM's acidotaxis-driven invasion.

- **Circadian Modulation:** Circadian rhythms modulate Meff and ion channel conductances (e.g., G_NHE, G_Ca_LEAK) every 3,600,000 ms (1 hour), with Meff decreasing hourly due to ROS-induced damage. Implemented in the model to model rhythmic mitochondrial degradation and frequency of proliferation over time.

All the 10 simulations for Stages B and C ran for 2,000 ms (40,000 steps, 0.05 ms timestep), initialized with $V_\square$ = –80 mV. The update loop is vectorized for efficiency. Parameters were chosen to account for limited computational resources.

### 3.2 Parameter Selection

Parameters were chosen for their roles in GBM dysregulation, with ranges informed by literature (*see Appendices A–B*). Each genome adjusts conductances (e.g., G_K_LEAK) and metabolic variables (e.g., Meff_init) via an update function to enable dynamic phenotype tuning in response to genetic variation within a controlled simulation framework. Conductances are sampled in log-space, while mutations for other parameters occur in real-space as log-space ensures biologically realistic scaling of channel conductances across orders of magnitude, while real-space is appropriate for bounded metabolic traits.

In Stage A, only Meff is varied (0.3 to 1.0), while all other parameters are fixed at healthy baselines to isolate the causal impact of mitochondrial dysfunction on bioelectric stability. In Stage B, Meff is fixed at 1.0, with all other parameters evolved. In Stage C, all parameters, including Meff, are evolved. Table 1 lists the evolvable parameters used in the simulation. Note that we use the union of the healthy and unhealthy ranges provided in literature as our effective range.

**Table 1. Evolvable Parameters in the Simulation**

| Parameter | Range | Description | Reference |
|---|---|---|---|
| G_K_LEAK S (cm⁻²) | $1 \times 10^{-9} - 5 \times 10^{-8}$ / $3 \times 10^{-4} - 1 \times 10^{-3}$ | Potassium leak conductance | (Brandalise et al., 2020; Takayasu et al., 2020) / (Zhong et al., 2023; Zhou et al., 2021) |
| G_NA_VG S (cm⁻²) | $5 \times 10^{-9} - 5 \times 10^{-8}$ / $2 \times 10^{-5} - 1 \times 10^{-4}$ | Voltage-gated sodium conductance | (Giammello et al., 2024; Molenaar, 2011) / (Lezmy et al., 2021; McNeill et al., 2021) |
| G_GAP (S cm⁻²) | $1 \times 10^{-5} - 1 \times 10^{-4}$ / $1 \times 10^{-4} - 5 \times 10^{-4}$ | Gap junction conductance | (McCutcheon & Spray, 2021) / (Cotrina et al., 1998; Ma et al., 2015) |
| Meff_init (%) | $0.3 - 1.0$ | Initial mitochondrial efficiency | Build-in value |
| PUMP_MAX (μA cm⁻²) | 0.3–1.0 / 0.5–1.5 | Na⁺/K⁺-ATPase pump rate | (Agrawal et al., 2023; Chen et al., 2014) / (Behbood et al., 2024; Yao et al., 2018) |

*3.3   Evolutionary Process*

Stage A uses a parameter sweep of Meff and does not involve evolution to directly assess the causal impact of metabolic efficiency on bioelectric stability without confounding factors. Stages B and C use a GA-based evolutionary loop, with Stage B fixing Meff and incorporating MAP-Elites for quality-diversity search to explore a broad and diverse range of stable and unstable phenotypes beyond metabolic variation, capturing resilience and vulnerability within the bioelectric landscape. In these stages, the GA evolves 96 tissue simulations over 10 generations using 15 CPU cores via Python's multiprocessing.Pool. Each simulation's genome comprises the parameters from Table 1. The process follows this loop (illustrated in Figure 2):

- **Initialization:** Random genomes, with Meff initialized homogeneously (no lesion unless stated) to isolate the effects of evolved parameters without introducing spatial heterogeneity or bias at the outset.
- **Selection:** Top 50% (48 individuals) survive via tournament selection. Tournament selection balances selection pressure and genetic diversity, preventing premature convergence
- **Reproduction:** Offspring use uniform crossover (50% chance per parameter) and Gaussian mutation (30% chance, σ=0.2 of range). Uniform crossover promotes exploration across all traits; Gaussian mutation introduces gradual variation without destabilizing the system
- **Immigration:** 20% (19 individuals) are replaced with random genomes per generation to maintain diversity (Eiben & Smith, 2015). To avoid genetic stagnation and ensure continued discovery of novel solutions.

PNG outputs and metrics are saved per generation; time-evolution snapshots are logged at the end of each generation for the best genome.

*3.4   Fitness Function and Evaluation Metrics*

The fitness function applies to evolutionary stages (B and C); Stage A is evaluated through V□ profiles and depolarization fraction. For Stages B and C, the fitness function targets GBM-like states:

$$Fitness \ = \ 0.1 \times CLIP \ Score \ + \ 0.35 \times Depol \ Fraction - 0.35 \times V\square \ RMSE \ + \ 0.2 \times V\square \ Entropy - M_{eff} \ Penalty - Overshoot \ Penalty$$

This fitness function was heuristically designed for this thesis to describe electrophysiological, morphological, and semantic aspects of GBM-like bioelectric states. While no single metric suffices to define malignancy, this weighted formula integrates: (i) a CLIP-based perceptual score aligned with human-labeled tumor morphology, (ii) depolarization fraction, (iii) distance from healthy V□ baseline (RMSE), (iv) spatial entropy, and penalties for excessive depolarization or low Meff. Such composite objectives are standard in evolutionary simulation studies (Bongard & Lipson, 2004; Clune et al., 2013). We use the ViT-B/32 CLIP model (Radford et al., 2021) on the CPU to compute similarity between V□ heatmaps and a "malignant glioblastoma with bright depolarized clusters" prompt to provide a high-level phenotypic fitness signal aligned with human-interpretable tumor morphology. Depolarization fraction (DF) counts cells with V□ > -40 mV, and an overshoot penalty applies if V□ exceeds -5 mV in over 20% of cells to discourage non-physiological overexcitation and maintain biological plausibility. Intermediate metrics are logged for tracking, while the final fitness score drives selection.

*3.5 Validation and Analysis*

Simulations are validated against known GBM electrophysiological traits: $V\square >$ –40 mV, elevated extracellular $Na^+/Ca^{2+}$, and spatial clustering (Cenciarini et al., 2019; Feher et al., 2022). A malignant-like state requires DF > 0.3. Clustering is quantified via spatial autocorrelation to capture non-random spatial organization characteristic of tumor microdomains. Outputs are visualized using Matplotlib's "coolwarm" colormap.

*3.6 Experimental Design*

To address our research questions, we conducted three experimental stages, each targeting a specific aspect of bioelectric instability in GBM:

- **Stage A (Causal Induction)**: We performed a controlled sweep of mitochondrial efficiency (Meff) from 0.3 to 1.0 in increments of 0.1, keeping all other parameters fixed at healthy baseline levels. Simulations ran for 3,000 ms on a 64×64 grid (60,000 steps, 0.05 ms timestep), tracking $V\square$, ATP, Psi_m, and ROS at central and peripheral cells. We measured the depolarization fraction (cells with $V\square > -40$ mV), aiming for a threshold of ≥40%, to test whether metabolic inefficiency alone induces depolarization and bioelectric instability.

- **Stage B (Resilience Mapping):** Simulations ran for 2,000 ms on a 32x32 grid 40,000 steps to balance resolution with runtime feasibility. With Meff fixed at 1.0 (healthy baseline), we applied the MAP-Elites algorithm (Mouret & Clune, 2015) to explore diverse bioelectric configurations. A CLIP-based diffusion model proposed new genomes by embedding similarity to high-performing $V\square$ heatmaps, acting as a guided mutation operator in addition to all other fitness score metrics.

- **Stage C (Emergent Tumor States):** Simulations ran for 2,000 ms on a 32×32 grid (40,000 steps, 0.05 ms timestep) to balance resolution with runtime feasibility. Full genome evolution, including Meff, was enabled to discover tumor-like attractor states under multi-objective selection. This tests the spontaneous emergence of GBM-like phenotypes. Parameter ranges included: G_K_LEAK (1e-9 to 1e-3), G_NA_VG (5e-9 to 1e-4), G_GAP (1e-5 to 2e-4), Meff_init (0.3 to 1.0), and PUMP_MAX (0.3 to 1.2). A 20% immigration rate was applied.

Table 2. Overview of Experimental Design Across Three Stages

| Stage | Method | Evolving Parameters | Goal |
|-------|--------|---------------------|------|
| A | Meff Sweep | Only Meff | Causal test of depolarization (RQ1) |
| B | MAP-Elites + Diffusion | All except Meff | Identify resilient/vulnerable states (RQ2) |
| C | Full GA | All (incl. Meff) | Evolve tumor-like attractors (RQ3) |

All experiments used GBM-adjacent initial conditions or in-between values (Appendices A–B), as fully healthy parameterizations resisted depolarization in 24-hour simulation windows due to GPU constraints. Table 2 summarizes the setup for each stage, and Figure 1 provides a visual overview of the three-stage experimental pipeline.

## 4   Results

This section presents the outcomes of our three-stage experimental design. Stage A examines the causal role of mitochondrial dysfunction in inducing depolarization (RQ1). Stage B maps resilient and vulnerable bioelectric states (RQ2), and Stage C explores the emergence of tumor-like attractors through full genome evolution (RQ3).

### 4.1   Stage A: Causal Induction of Depolarization via Metabolic Inefficiency

Stage A tested whether mitochondrial inefficiency alone—operationalized as a reduction in mitochondrial efficiency (Meff) from 1.0 to 0.3—was sufficient to trigger depolarization in glial tissue. Simulations were run on a 64×64 grid for 3,000 ms using a tissue-scale numerical model of ion dynamics, membrane voltage ($V\square$), ATP production, and reactive oxygen species (ROS). The 3,000 ms window was selected to capture early, acute bioelectric transitions without invoking long-term compensatory mechanisms, while the grid size approximates a mesoscopic tissue patch at single-cell resolution. A cell was classified as depolarized if $V\square$ exceeded −40 mV, consistent with electrophysiological thresholds reported in glioblastoma.

The Meff range (1.0 to 0.3) was chosen to simulate a spectrum from normal oxidative phosphorylation to severe mitochondrial dysfunction, capturing the metabolic collapse and oxidative stress characteristic of the Warburg effect in GBM (Bost et al., 2015; Gatenby & Gillies, 2004). As Meff decreased, the fraction of depolarized cells increased markedly (Figure 4.1). At Meff = 1.0 (healthy baseline), nearly all cells maintained a hyperpolarized state (mean $V\square \approx$ −80 mV), and the depolarization fraction remained below 1%. At Meff = 0.6, this fraction crossed the 40% threshold, reaching ~42%, suggesting a bifurcation point in the system's dynamics. At Meff = 0.3, 98% of cells were depolarized (mean $V\square \approx$ −35 mV). Further analysis needs to be done to check if depolarized regions actually form non-random patches that are consistent with GBM's tumor microdomains.

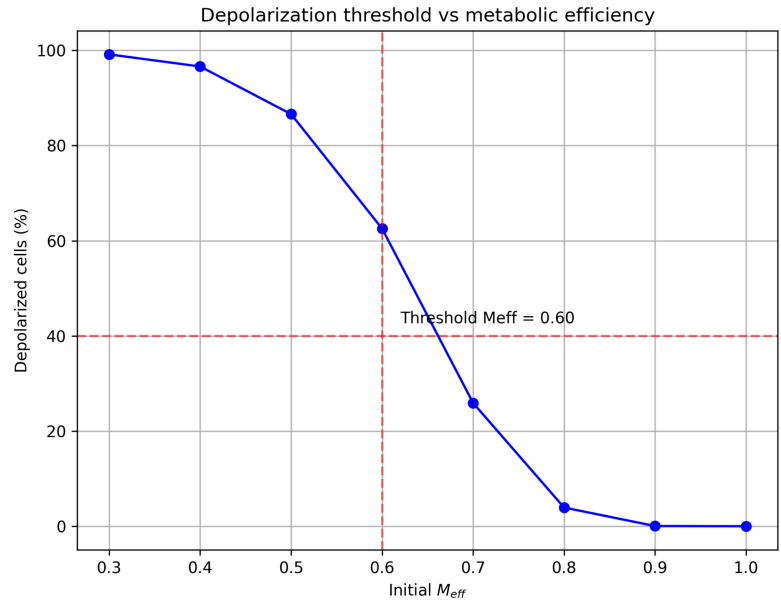

Figure 4. Depolarization Fraction vs. Meff.
Directly quantifies the threshold for GBM-like instability, addressing RQ1's focus on causality. Horizontal and vertical red dashed lines mark the 40% depolarization threshold and the bifurcation point at Meff = 0.6.

Spatial membrane potential distributions further illustrated this transition ([Figure 4](#)). At Meff = 1.0, the tissue was uniformly hyperpolarized. At Meff = 0.6, scattered patches of depolarization emerged, signaling the onset of instability. At Meff = 0.3, large-scale depolarized clusters dominated the grid, with V▯ values nearing −30 mV—hallmarks of GBM-like electrophysiological disruption.

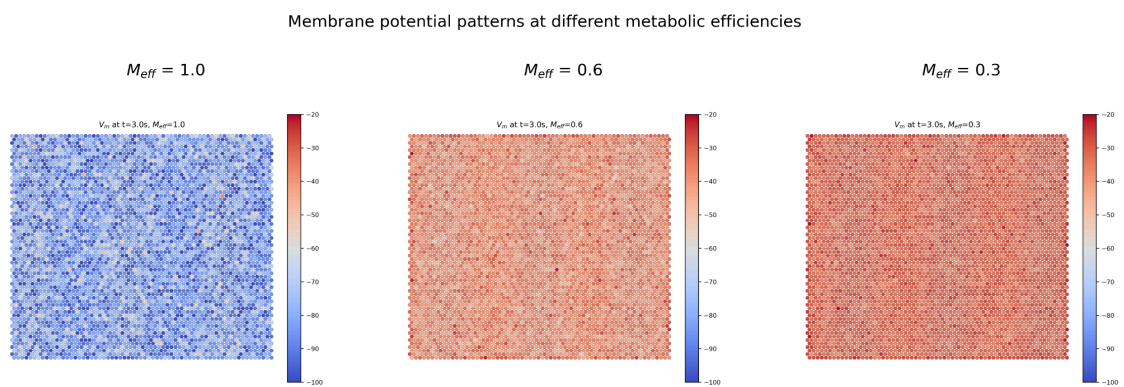

Figure 5. V▯ Patterns at Different Meff Values.
Three-panel heatmap showing final V▯ distributions for Meff = 1.0, 0.6, and 0.3 at t = 3,000 ms. Shows spatial heterogeneity to better understand how depolarization spreads in tissue.

To capture the temporal dynamics of this transition, time-course data were recorded at two representative locations: a central cell (H/2, W/2) and a peripheral cell (H/4, W/4) —center (32, 32) and peripheral (16, 16)—to capture spatial variability in bioelectric responses. Under severe mitochondrial inefficiency with $M_eff = 0.3$ (Figure 6) both the central (32, 32) and peripheral (16, 16) cells began near −80 mV and drifted steadily toward depolarisation throughout the 3 s run. The central cell crossed the −40 mV threshold after roughly 2.3 s and finished at −33 mV, while the peripheral cell lagged slightly and ended near −36 mV, indicating only modest spatial heterogeneity. Cytosolic ATP, initially just below 1 mM, collapsed to less than 10 μM within the first 100 ms and remained essentially exhausted, whereas the mitochondrial membrane potential softened only marginally from −175 mV to −171 mV, pointing to a partial—but not catastrophic—loss of the proton-motive force. Reactive-oxygen species accumulated almost linearly, reaching about $2.6 \times 10^3$ arbitrary units by 3 s and signalling intense oxidative stress, driving DNA damage and tumor progression (Liu et al., 2022; Srinivas et al., 2018; Yang et al., 2023).. In contrast, when mitochondrial efficiency was kept at its healthy value $M_eff = 1.0$ (Figure 7) both recording sites stayed hyperpolarised (between −95 mV and −75 mV), the mitochondrial potential remained stable, ATP did not crash, and ROS rose only slowly, confirming that the dramatic depolarisation and redox surge observed at $M_eff = 0.3$ are direct consequences of metabolic failure rather than numerical artefact or boundary effects.

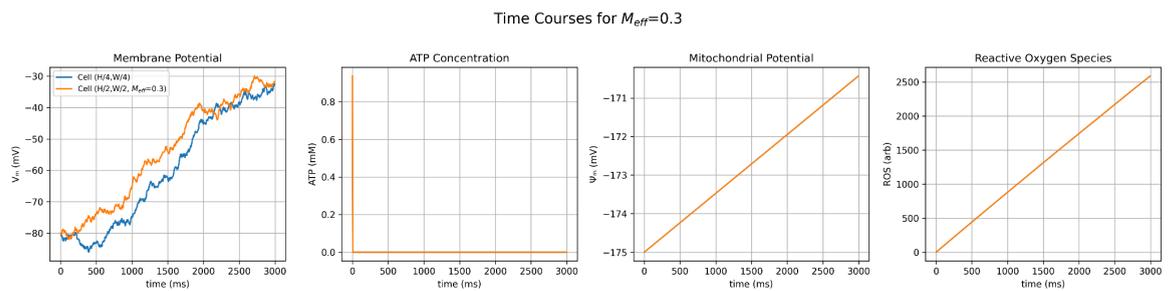

Figure 6. Time-Course for Meff = 0.3.
Four-panel plot showing V☐, ATP, Psi_m, and ROS over 3,000 ms for a central (H/2, W/2) and peripheral (H/4, W/4) cell at Meff = 0.3. Shows the progression of instability.

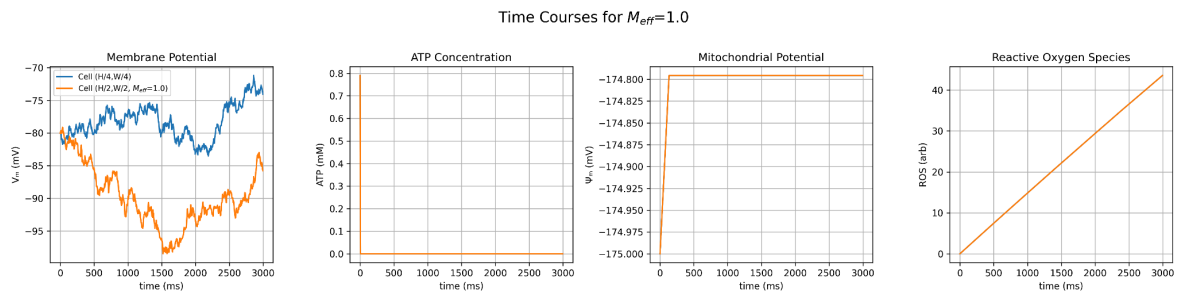

Figure 7. Time-Course for Meff = 1.0
Four-panel plot showing V☐, ATP, Psi_m, and ROS over 3,000 ms for a central (H/2, W/2) and peripheral (H/4, W/4) cell at Meff = 1.0.

A timelapse of V☐ evolution across all Meff values (Supplementary Figure S1) highlights the progressive spread of depolarization as Meff decreases, with lower Meff values showing faster and more widespread depolarization within the 3,000 ms simulation window.

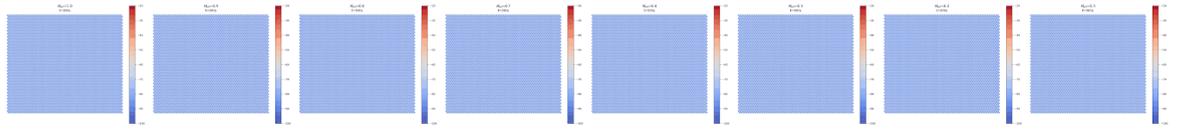

Supplementary Figure S1. GIF showing V▯ evolution across all Meff values (1.0 to 0.3, left to right) over a duration of 3,000 ms. Depolarization onset is earlier and more widespread for lower Meff values. Visualizes the dynamic spread of depolarization, showing a comprehensive view of the process over time. >Enlarge<

Additional whole-grid metrics (Figure 8) reinforce the central role of mitochondrial efficiency in setting the tissue's bioelectric state. As the initial $M_e$ff was lowered from 1.0 to 0.3 the mean membrane potential drifted monotonically from $-82$ mV to $-32$ mV, with a pronounced elbow between 0.8 and 0.7 that mirrors the bifurcation seen in the single-cell traces. The most-depolarised cell in the lattice plateaued near $-20$ mV for $M_e$ff $\leq 0.5$, indicating that once oxidative capacity falls below this threshold further energetic loss no longer yields additional positive swing. In contrast, the peak extracellular potassium concentration remained fixed at $\approx 20$ mM across the entire $M_e$ff range, implying that $[K^+]_e$ saturates early and is not the variable that drives the progressive drop in mean V▯. Collectively these trends show that metabolic failure alone can push an otherwise healthy glial network into the depolarised voltage regime characteristic of GBM, even without any further rise in extracellular $K^+$ or other secondary insults.

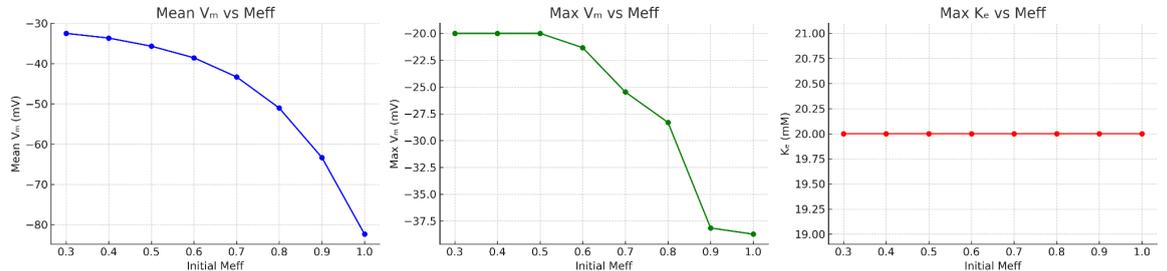

Figure 8. Electrophysiological metrics as a function of initial mitochondrial efficiency.
Left: tissue-averaged membrane potential after 3 s.
Centre: maximum membrane potential observed in the grid.
Right: maximum extracellular potassium concentration.
Lower $M_e$ff values produce a sigmoidal loss of polarisation (left) and a ceiling of about $-20$ mV on the most-depolarised cells (centre), while the peak $[K^+]_e$ quickly saturates at 20 mM and is essentially independent of $M_e$ff (right).

### 4.2  Stage B: Resilience Mapping of Bioelectric States

Stage B evolved parameters (G_K_LEAK, G_NA_VG, G_GAP, PUMP_MAX) with fixed Meff = 1.0 using a genetic algorithm (96 genomes, 10 generations, 32×32 grid, 2,000 ms, 0.05 ms timestep, 20% immigration rate), targeting stability via a fitness function penalizing V▯ deviation from -80 mV. The depolarization fraction (V▯ > -40 mV) remained 0.0 across generations. Best fitness rose from 0.0199 (Gen 0) to 0.0421 (Gen 9[1]), plateauing after Gen 2

---

[1] zero indexed

(0.0394), with average fitness improving from -0.1333 to 0.0060 and worst fitness ranging from -0.1593 to -0.1557 (Figure 9). V□ RMSE, the primary optimization driver, decreased from 21.26 mV to 17.95 mV, reflecting a mean V□ shift from -58.7 mV to -62.1 mV (v_rmse_n: 0.3543 to 0.2986). The CLIP score (0.6267 to 0.6316) and V□ entropy (0.4052 to 0.4183) remained relatively stable, indicating no real effect of morphology and electrical homeostasis on this optimization case.

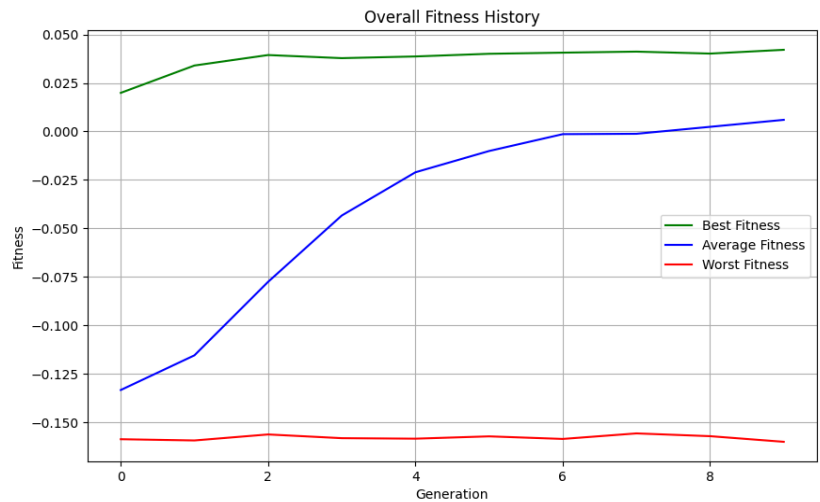

Figure 9. Fitness history of the population in stage B across 10 generations, showing best, average, and worst fitness scores.

Multiple parameter combinations yielded near-optimal fitness. G_K_LEAK increased from 0.00083 to 0.001 S cm⁻² by Generation 2, G_NA_VG varied from 2.04e−7 to 4.87e−5 S cm⁻², G_GAP ranged from 1.09e−5 to 2.46e−5 S cm⁻², and PUMP_MAX fluctuated between 0.31 and 1.2 (Figure 9).

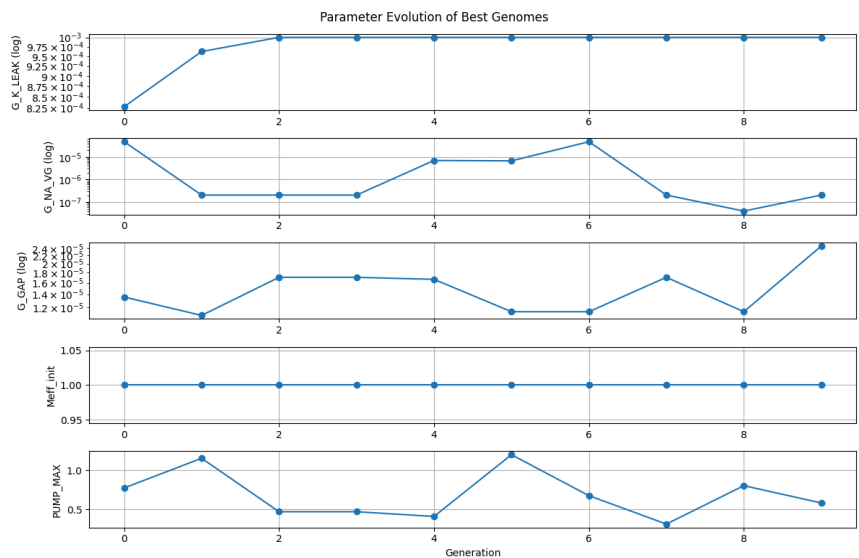

Figure 10. Parameter evolution plots (G_K_LEAK, G_NA_VG, G_GAP, PUMP_MAX) across 10 generations for the best genome per generation, using a logarithmic scale for conductances.

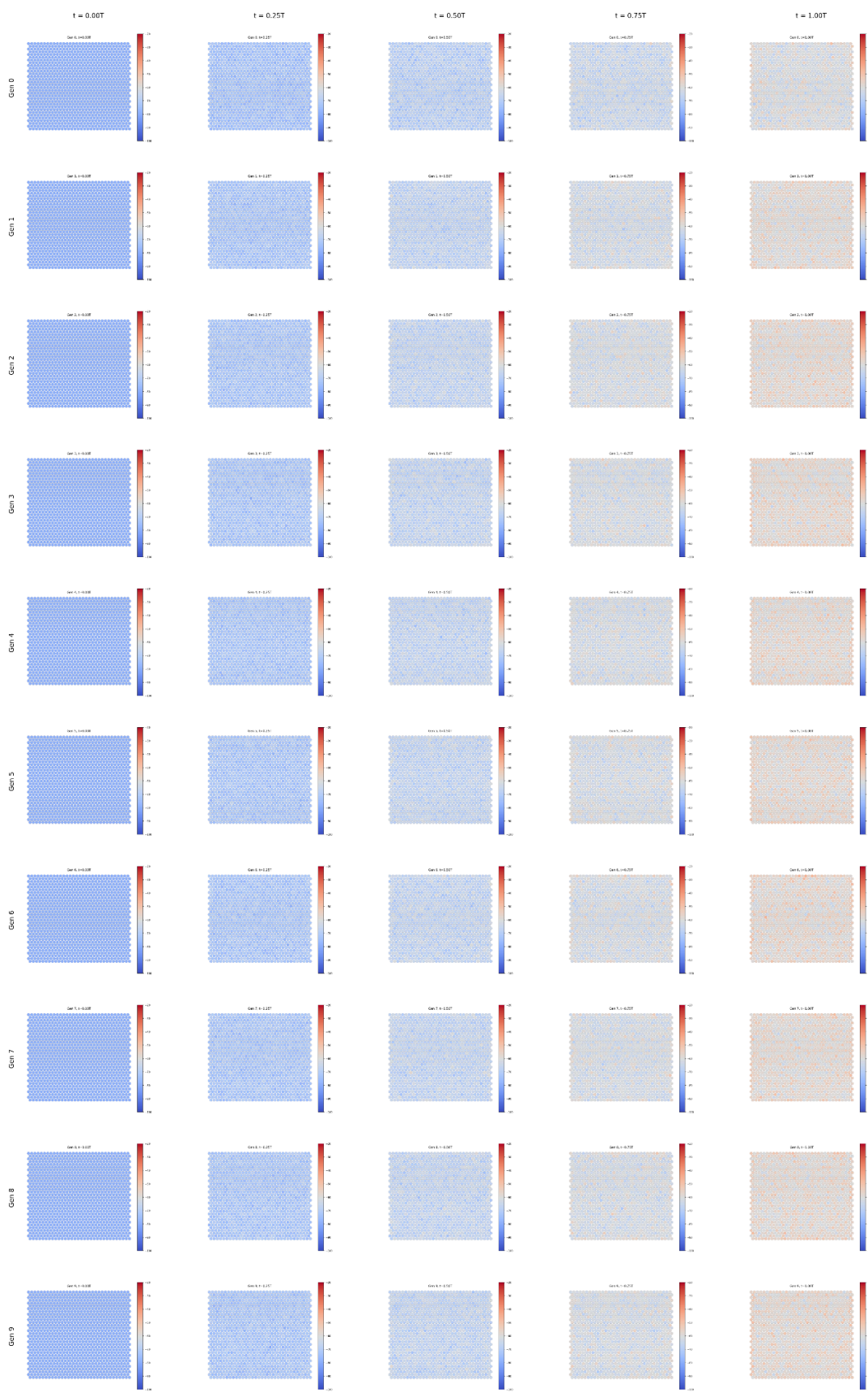

Figure 11. Five-panel heatmap showing V□ distributions for the best genome in Generation 9 at t = 0, 500, 1,000, 1,500, and 2,000 ms. >Enlarge<

The final V☐ distribution in Generation 9 showed a uniformly polarized pattern, with no depolarized clusters. Time-evolution snapshots at t = 0, 500, 1,000, 1,500, and 2,000 ms confirmed membrane potential stability (Figure 4.9). These outcomes confirm that even under genetic variation of major conductances, the bioelectric state remained stable and non-tumorigenic when Meff was fixed at healthy levels.

### 4.3  Stage C: Emergence of Tumor-Like Attractors via Full Genome Evolution

Stage C evolved all parameters, including Meff, using a genetic algorithm with a population of 96 genomes over 10 generations on a 32×32 grid. Each simulation ran for 2,000 ms (40,000 steps, 0.05 ms timestep) with a 20% immigration rate. The algorithm achieved a peak depolarization fraction of 0.9746 and a best fitness score of 0.4043 by Generation 9, with average fitness increasing from 0.0280 to 0.3093 across generations (Figure 12).

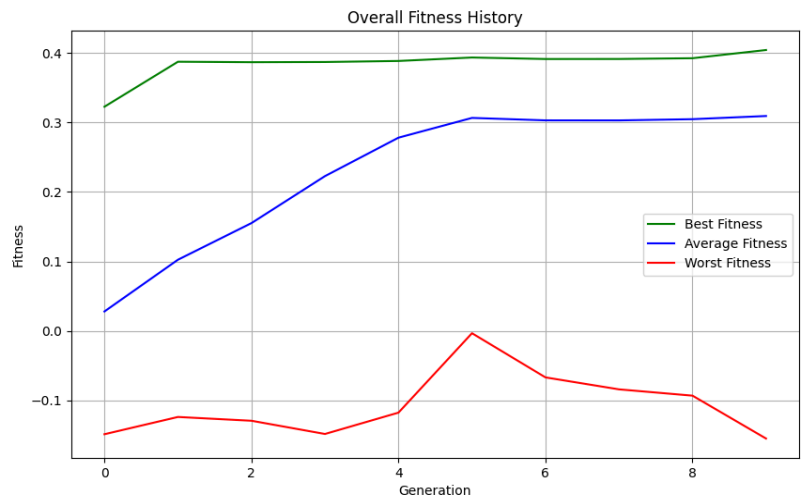

Figure 12. Fitness history, plotting best, average, and worst fitness scores per generation.

The depolarization fraction (fraction of cells with V☐ > –40 mV) rose from 0.6904 in Generation 0 to 0.9746 in Generation 9. Mean V☐, derived from V☐ RMSE, shifted from approximately –52 mV (v_rmse_n = 0.0398) in Generation 0 to –29.8 mV (v_rmse_n = 0.0702) in Generation 9. *Fitness components such as CLIP score and entropy were tracked internally but are not shown in this figure.* Figure 12 shows the overall fitness history only, including best, average, and worst fitness per generation.

Evolved parameters in the best genomes included Meff_init decreasing from 0.3867 in Generation 0 to 0.3 by Generation 9, G_K_LEAK ranging from 0.0007 to 0.0009 S cm⁻², G_NA_VG increasing to 5e⁻⁵ S cm⁻² in Generation 8, G_GAP decreasing from 0.00019 to 3e⁻⁵ S cm⁻², and PUMP_MAX varying between 0.5834 and 1.0 (Figure 13).

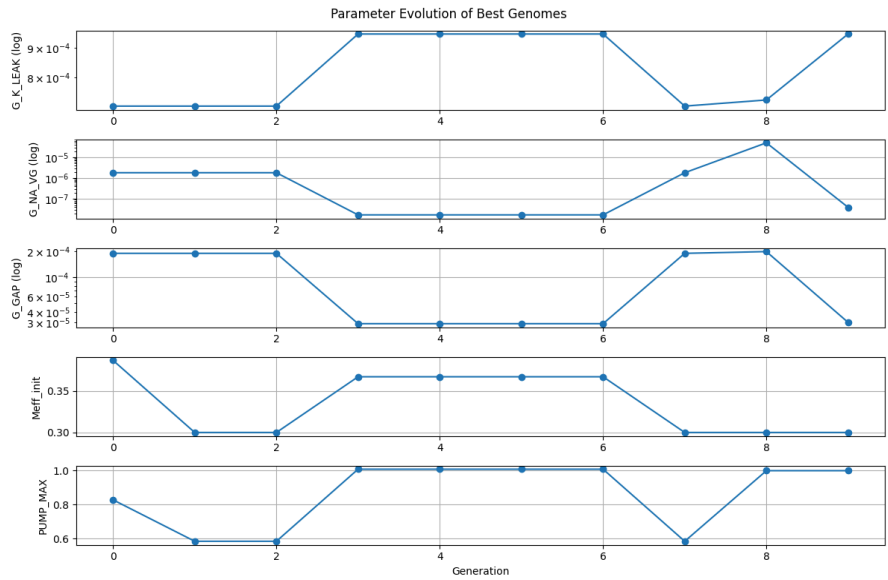

Figure 13. Line plots showing the values of evolved parameters (Meff_init, G_K_LEAK, G_NA_VG, G_GAP, PUMP_MAX) in the best genome for stage C across 10 generations, with conductance parameters on a logarithmic scale.

The final V☐ distribution in Generation 9 showed widespread depolarization across the grid, with values concentrated around -29.8 mV. Time evolution snapshots at t = 0, 500, 1,000, 1,500, and 2,000 ms for Generation 9 showed the spread of depolarization, with clustering evident by 1,000 ms (Figure 14).

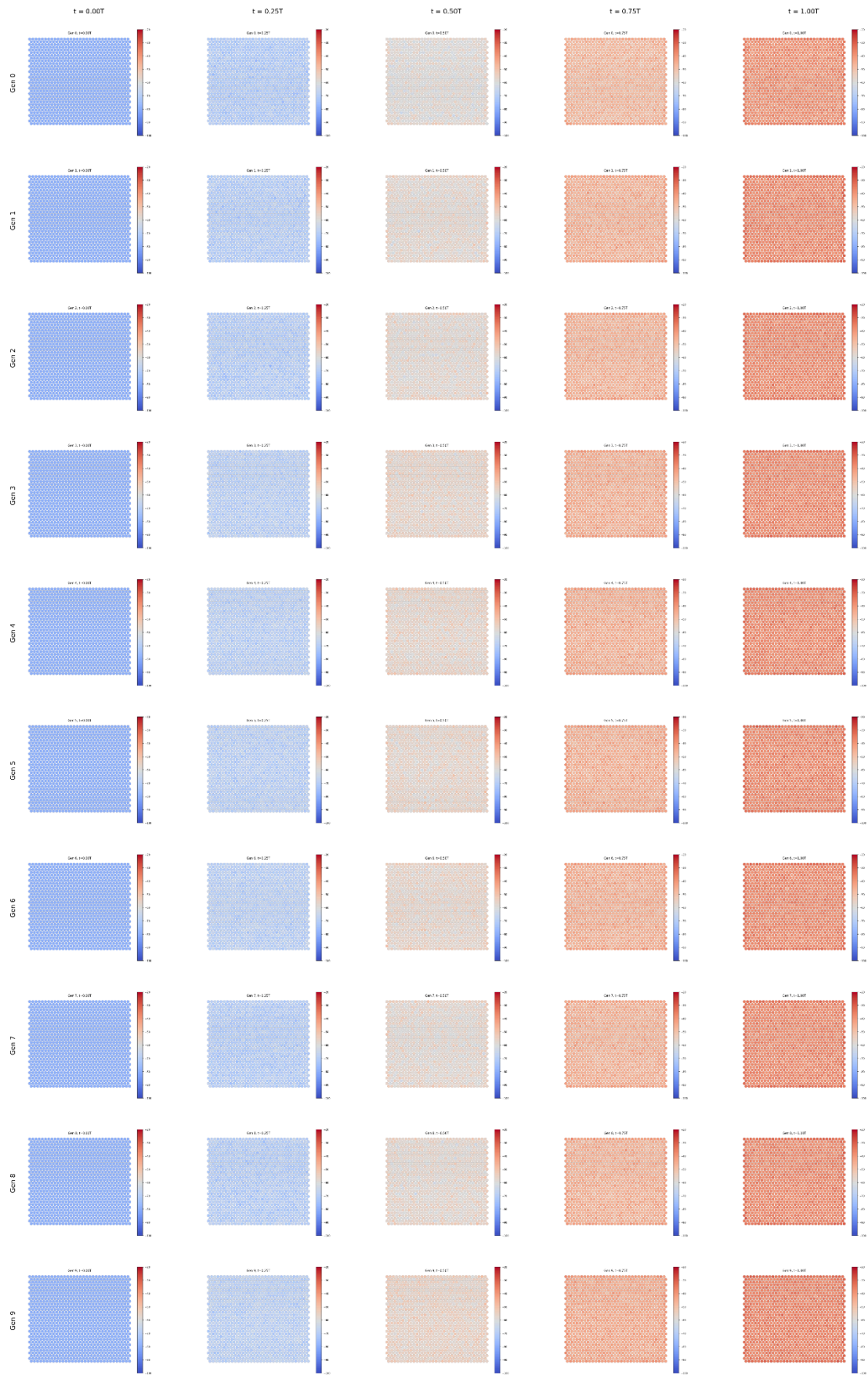

Figure 14. Five-panel heatmap showing V□ distributions at t = 0, 500, 1,000, 1,500, and 2,000 ms for the best genome in Generation 9. >Enlarge<

## 5    Discussion

This study investigated the bioelectric transition of glial cells into glioblastoma (GBM)-like states using a three-stage computational framework. Stage A established the causal role of mitochondrial dysfunction in inducing depolarization (RQ1), while Stages B and C were designed to map bioelectric resilience (RQ2) and explore the emergence of tumor-like attractors (RQ3). Below, we interpret the results of Stage A, B and C, consider their biological relevance, and discuss model assumptions.

### 5.1    Stage A: Mitochondrial Dysfunction as a Driver of Bioelectric Instability

Stage A confirmed that mitochondrial efficiency is a single-parameter switch for glial bioelectric stability. Decreasing $M_e$ff from the healthy value of 1.0 to 0.3 produced a monotonic, sigmoidal loss of polarisation across the $64 \times 64$ lattice. A tipping region appeared at $M_e$ff $\approx 0.7 - 0.6$: once efficiency fell into this band the fraction of cells with $V\square > -40$ mV jumped past 40 %, signalling a system-level bifurcation. At the extreme impairment ($M_e$ff $= 0.3$) 98 % of cells were depolarised and the tissue-average membrane potential settled near $-33$ mV, while the most-depolarised sites plateaued around $-20$ mV. Concomitantly, cytosolic ATP collapsed from $\sim 0.9$ mM to $<10$ μM within the first 100 ms, mitochondrial $\Psi\square$ weakened only slightly ($-175 \rightarrow -171$ mV), and ROS rose linearly to $\sim 2.6 \times 10^3$ a.u. by 3 s. Extracellular $K^+$ spiked rapidly to its saturation ceiling of $\approx 20$ mM and then remained flat across all lower $M_e$ff values, indicating that continued depolarisation after the initial pump failure is driven chiefly by leak conductances rather than further $K^+$ accumulation. No additional genetic or micro-environmental perturbations were required, establishing $M_e$ff as a minimal sufficient driver of GBM-like electrophysiology in this model.

These simulated signatures map closely onto documented glioblastoma physiology. Membrane potentials in the $-30$ mV range foster proliferation and invasion by bringing voltage-gated $Na^+$, $Ca^{2+}$, and $K^+$ channels into their operational windows (Yang & Brackenbury 2013; Turner & Sontheimer 2014). The rapid, near-total ATP depletion mirrors the Warburg shift, where GBM cells abandon oxidative phosphorylation yet sustain growth through accelerated glycolysis under mitochondrial stress (Seyfried 2015). A fixed high $[K^+]_e$ around 20 mM echoes in-vivo reports of peritumoural $K^+$ plateaus that promote excitotoxicity and oedema (Olsen & Sontheimer 2008), while the thousand-fold ROS surge aligns with the oxidative-damage loop that drives genomic instability and tumour progression (Salazar-Ramiro et al. 2016). The sharp transition between $M_e$ff $\approx 0.7$ and 0.6 therefore marks a plausible bioelectric tipping point for malignant transformation. If verified experimentally, rescuing mitochondrial efficiency above this threshold—via metabolic modulators such as dichloroacetate or mito-targeted antioxidants (Michelakis et al. 2010)—could stabilise glial membrane potential and delay, or even prevent, GBM initiation. Thus, Stage A strengthens the view that bioelectric state is not merely a read-out but a causal regulator of oncogenic behaviour, augmenting genetic and epigenetic models of tumourigenesis and supporting broader efforts to treat cancer through electrophysiological and metabolic re-normalisation.

### 5.2    Stage B Resilience Mapping of Bioelectric States

Stage B mapped bioelectric resilience with fixed Meff = 1.0 (RQ2), shows that no parameter combination led to depolarization, with the depolarization fraction remaining at 0.0 across all generations. This shows bioelectric stability in the absence of mitochondrial dysfunction, contrasting with Stage A's findings, where reducing Meff below 0.6 caused significant depolarization. The mean $V\square$ in Stage B, ranging from

-58.7 mV to -62.1 mV, stayed well below the GBM threshold (-40 mV), showing protective role of healthy mitochondrial function in maintaining glial hyperpolarization.

The fitness function, driven primarily by V☐ RMSE, optimized configurations toward the GBM threshold (-40 mV), with the best fitness score plateauing early (by Generation 2) and remaining near 0.04. This fast convergence might suggests a strong selective pressure for stability, but the plateau indicates a limit to further optimization, possibly due to the fitness function's emphasis on minimizing V☐ deviation over other factors like spatial heterogeneity (e.g., V☐ entropy increased only slightly). The presence of multiple parameter combinations achieving near-optimal fitness possibly points to the existence of diverse bioelectric configurations that can maintain stability. For instance, G_K_LEAK consistently reached its upper bound, possibly enhancing potassium efflux to stabilize V☐, while G_NA_VG varied widely, suggesting sodium conductance was less important in this context. Moreover the range in G_GAP and PUMP_MAX further show that both gap junction coupling and pump activity can vary without compromising hyperpolarization, possibly suggesting that there is a degree of bioelectric redundancy in healthy states.

Although we did not fully explore the whole bioelectric state under Meff = 1.0 condition due to computational limitations, biologically, our limited results suggest that glial tissue with intact mitochondrial function can resist malignant transitions, even under varied ion channel and pump dynamics, supporting the hypothesis that bioelectric resilience depends heavily on metabolic health (Levin, 2014; Mathews et al., 2022; Boyle et al., 2022). The absence of depolarization aligns with experimental observations of healthy glia maintaining V☐ around -70 mV to -80 mV (Molenaar, 2011; Viola et al., 2024). However, the diversity in optimal configurations implies that resilience is not tied to a single "ideal" parameter set, which could complicate possible therapeutic strategies trying to reinforce bioelectric stability in pre-malignant states. Compared to Stage C, where evolving Meff led to GBM-like states, Stage B shows that mitochondrial efficiency is an important factor in bioelectric failure and stability. Future studies could explore whether introducing mild metabolic stress (e.g., Meff slightly below 1.0) or microenvironmental perturbations (e.g., hypoxia) might push these resilient states toward vulnerability, as hypothesized in the original design of Stage B.

### 5.3 Stage C Emergence of Tumor-Like Attractors

Stage C confirmed that tumor-like attractor states spontaneously emerge under evolutionary pressure, supporting RQ3. The genetic algorithm drove the population toward GBM-like phenotypes, achieving high depolarization and mean V☐ within GBM's characteristic range (-40 mV to -20 mV) by Generation 9. Evolved parameters mirrored GBM electrophysiology: Meff_init converged to a low value, below the threshold identified in Stage A, showing the important role of mitochondrial dysfunction in depolarization. Reduced G_K_LEAK and increased G_NA_VG reflected the "oncochannel" phenotype, enhancing sodium currents that drive proliferation and invasion (McCutcheon & Spray, 2021; Litan & Langhans, 2015), while decreased G_GAP aligned with impaired gap junction coupling in GBM, facilitating uncoordinated cell behavior (McCutcheon & Spray, 2021).

The high depolarization and spatial clustering of V☐ suggest that these tumor-like states are stable attractors in the bioelectric landscape, rather than transient states. This evolutionary convergence shows that GBM-like phenotypes may be energetically favorable under metabolic and bioelectric stress, challenging the view of tumors as purely chaotic systems (McCutcheon & Spray, 2021; Litan & Langhans, 2015) Instead, they may represent adaptive states optimized for survival and proliferation, consistent with theories of cancer as an atavistic reversion to simpler, unicellular-like behaviors (Davies & Lineweaver, 2011; Nikitin, 2022). The consistent GBM-like morphology across generations, as indicated by stable CLIP scores, suggests that depolarized clusters could be a diagnostic signature if validated experimentally.

Compared to Stage A, Stage C shows synergistic effects of multiple parameters evolving together, amplifying the depolarization driven by mitochondrial dysfunction and pushing the system further into a malignant regime. This might suggest that therapeutic strategies targeting only mitochondrial function may be insufficient; simultaneous modulation of ion channel activity (e.g., using ion channel blockers) or gap junction coupling could disrupt these attractors more effectively (Levin, 2014). The evolutionary stability of these states also raises questions about GBM's resistance to treatment, as such attractors may enable rapid adaptation to therapeutic pressures.

### 5.4   Limitations and Future Directions

While Stage A provides compelling evidence for mitochondrial-driven depolarization, the model simplifies certain aspects of GBM biology. Stage A used a 64×64 grid (4,096 cells) to achieve higher spatial resolution, but Stages B and C were constrained to a 32×32 grid (1,024 cells) due to the significant computational demands of the genetic algorithm, which required more iterations to capture evolutionary dynamics while exploring the parameter space. This reduction in grid size may limit the spatial complexity of bioelectric patterns in Stages B and C, potentially underestimating clustering or heterogeneity. Simulations were limited to 3,000 ms in Stage A and 10,000 ms in Stages B and C due to GPU memory and runtime constraints. While these durations were sufficient to capture acute transitions and short-term evolutionary behavior, they excluded slower physiological processes such as circadian modulation (active only in Stages B and C), cell cycle entry, or glial proliferation. Additionally, although the 64×64 grid in Stage A offered higher resolution, it still does not reflect the full three-dimensional complexity of brain tissue, where cell density and extracellular matrix interactions play a role. Also, the optimization function could be improved, both by choosing better component fitness scores and also better relative weight tuning. When increasing the penalty on lowering Meff, we reach higher Meff solutions proving that there is more potential for exploration here.

Additionally, the model omits immune responses and angiogenesis, which are crucial in GBM progression in vivo. Several modeling simplifications were made to isolate the causal effects of mitochondrial dysfunction. First, glial cells were treated as electrically and metabolically homogeneous, omitting stochastic variability in ion channel expression or mitochondrial content observed in vivo (Lorimer, 2024; Pridham et al., 2022; Tatla et al., 2021). Second, gap junctions were modeled using static conductance values, without dynamic remodeling, although connexin expression is known to vary in GBM. Third, all cells started with identical resting conditions, excluding potential epigenetic or microenvironmental priming that might affect bioelectric responses. Fourth, despite meeting the V□ threshold for division (−38 mV > −40 mV), cell division and cell migration were not observed at Meff = 0.3 due to the short 3,000 ms timescale. Another limitation is the fact that due to computational limitations we could not evolve all the possible parameters which might drastically change the modeled behavior of cells under certain biological conditions.

Moreover, ROS and ATP dynamics were implemented using simplified rules, and their concentrations were expressed in relative model units not anchored to experimentally calibrated kinetics. Consequently, the long-term stability or reversibility of depolarized states remains unknown. Extending runtime via hybrid GPU–CPU scheduling or reducing spatial resolution could make longer simulations feasible. Alternatively, cloud computing could enable the parallel simulation of larger populations or extended biological periods. Incorporating stochastic dynamics, transcriptomic state transitions, or 3D multicellular architectures could further improve the biological realism of the model.

The choice of Meff as the sole variable in Stage A, while effective for isolating causality, overlooks the multifactorial nature of GBM. In vivo, tumor initiation likely involves complex interactions between metabolic, electrophysiological, immunological, and genetic mechanisms. Stages B and C likely addressed this by exploring broader parameter spaces, but their incomplete results limit our ability to draw definitive conclusions. Future iterations of this study should prioritize completing these stages, focusing on how resilience (Stage B) and evolutionary dynamics (Stage C) interact with mitochondrial dysfunction. Experimentally, validating these findings in vitro using GBM cell lines with induced mitochondrial stress could confirm the Meff threshold and its bioelectric consequences.

Finally, the therapeutic implications of this work are promising but speculative. The Meff threshold suggests a window for early intervention, but clinical translation requires identifying measurable proxies for Meff in patients, such as mitochondrial membrane potential or ATP/ADP ratios. More broadly, metabolic or electrical reprogramming may offer a non-genetic strategy for delaying or reversing malignant transitions in the brain, though additional research is needed to develop clinically viable protocols.

## 6    Conclusion

This study suggests mitochondrial dysfunction as a potential driver of bioelectric instability in glial cells, with an Meff threshold of 0.6 marking GBM-like depolarization. These insights show early GBM mechanisms and suggest bioelectric and metabolic therapeutic targets. Anticipated outcomes from Stages B and C point to resilience and evolutionary dynamics as critical factors in GBM's bioelectric landscape. Bridging these computational findings with experimental and clinical validation is essential to confirm correctness of the model and further determine GBM treatment strategies.

Appendix A. Healthy Glia Cell Model Parameters

| Code constant | Unit (model) | What it is / what it does | Value used | Healthy range | Key refs |
|---|---|---|---|---|---|
| **F** | C mol⁻¹ | Faraday constant – charge per mole e⁻ | 96 485.332 9 | N/A (physical) | Constant physics values |
| **R** | J mol⁻¹ K⁻¹ | Gas constant | 8.314 462 618 | N/A | Constant physics values |
| **T** | K | Physiological T (37 °C) | 310.15 | 309-311 (normothermia) | Constant physics values |
| **RT_F** | mV | Thermal voltage (RT/F) | 26.73 | N/A | Constant physics values |
| **C_M** | µF cm⁻² | Specific membrane capacitance of astroglial membrane | 1.0 | ≈1.0 ±0.1 | (Savtchenko et al., 2018; McNeill et al., 2021) |
| **G_K_LEAK** | S cm⁻² | Passive (ohmic) K⁺ leak channels (Kir4.1 + K2P mix) | $6 \times 10^{-4}$ | $3 \times 10^{-4} - 1 \times 10^{-3}$ | (Zhong et al., 2023; Zhou et al., 2021) |
| **G_NA_LEAK** | S cm⁻² | TTX-resistant Na⁺ background (NALCN/Na$_\chi$) | $1 \times 10^{-4}$ | $8 \times 10^{-5} - 3 \times 10^{-4}$ | (McNeill et al., 2021; Ren, 2011) |
| **G_NA_VG** | S cm⁻² | Sparse fast Na⁺ VG channels in perisynaptic processes | $5 \times 10^{-5}$ | $2 \times 10^{-5} - 1 \times 10^{-4}$ | (Lezmy et al., 2021; McNeill et al., 2021) |
| **G_Ca_LEAK** | S cm⁻² | Ca²⁺ leak across PM (TRP/Orai background) | $5 \times 10^{-6}$ | $2 \times 10^{-6} - 1 \times 10^{-5}$ | (Lezmy et al., 2021; Tewari & Majumdar, 2012) |
| **G_Cl_LEAK** | S cm⁻² | Cl⁻ leak / VRAC baseline | $5 \times 10^{-5}$ | $\leq 5 \times 10^{-5}$ (alt. $1 \times 10^{-4}$–$5 \times 10^{-4}$) | (Lüscher et al., 2019; Raut et al., 2024) |

| **PUMP_MAX** | μA cm$^{-2}$ | Na$^+$/K$^+$-ATPase max turnover current density | 1.0 | 0.5–1.5 | (Behbood et al., 2024; Yao et al., 2018) |
|---|---|---|---|---|---|
| **I_PMCA_MAX** | μA cm$^{-2}$ | PMCA Ca$^{2+}$ pump saturation current | 2.55 × 10$^0$ | 1 × 10$^{-8}$ – 5 × 10$^{-7}$ | (Tewari & Majumdar, 2012) |
| **G_GAP** | S cm$^{-2}$ | Connexin-43/30 gap-junction sheet conductance | 2 × 10$^{-4}$ | 1 × 10$^{-4}$ – 5 × 10$^{-4}$ | (Cotrina et al., 1998; Ma et al., 2015) |
| **ATP_BASE** | mM | Resting cytosolic ATP | 1.3 | 1.2–1.4 | (Köhler et al., 2020; Lerchundi et al., 2020) |
| **ATP_KM** | mM | Pump Michaelis constant | 0.6 | 0.5–0.7 | (Behbood et al., 2024; Tewari & Majumdar, 2012) |
| **ATP_REGEN_RATE** | mM ms$^{-1}$ | Basal mitochondrial/reglycolytic ATP regen | 2 × 10$^{-3}$ | 1–3 × 10$^{-3}$ | (Lezmy et al., 2021; Sriram et al., 2024) |
| **ATP_PER_UA_MS** | mM /(μA·ms) | ATP cost per ionic current | 31 | Derived | ATP_PER_UA_MS = $\frac{10^3 A_{cell}}{F V_{cell}}$ |
| **AREA_CM2** | cm$^2$ | 20 μm astrocyte surface area | 1.26 × 10$^{-5}$ | 1.1–1.4 × 10$^{-5}$ | (Savtchenko et al., 2018) |
| **VOL_CM3** | cm$^3$ | 20 μm cell volume | 4.19 × 10$^{-9}$ | 4–4.5 × 10$^{-9}$ | (Savtchenko et al., 2018) |
| **K_E** | mM | Typical brain ECF [K$^+$] | 3.0 | 2.5–3.5 | (Berret et al., 2014; Paulson & Newman, 1987) |
| **Na_E** | mM | ECF [Na$^+$] | 145 | 140–150 | (Hagiwara et al., 2021; Noda & Hiyama, 2014) |
| **Cl_E** | mM | ECF [Cl$^-$] | 120 | 115–125 | (Raut et al., 2024) |
| **Ca_E** | mM | ECF [Ca$^{2+}$] | 1.3 | 1.2–1.4 | (Inglebert et al., 2020; Lu et al., 2010) |

| **G_NHE** | mM ms$^{-1}$ per $\Delta[H^+]$ | Na$^+$/H$^+$ exchanger proton extrusion rate | $5 \times 10^{-5}$ | $3–7 \times 10^{-5}$ | (Lezmy et al., 2021; Lu et al., 2010) |
|---|---|---|---|---|---|
| **KF_CO2 / KR_CO2** | mM$^{-1}$ ms$^{-1}$ / ms$^{-1}$ | CO$_2$ hydration / dehydration kinetics | $5 \times 10^{-2}$ / $2 \times 10^{-2}$ | $4–6 \times 10^{-2}$ / $1–3 \times 10^{-2}$ | (Cadogan et al., 2014; Endeward et al., 2017) |
| **GLY_MAX** | mM ms$^{-1}$ | Max glycolytic ATP prod. | $1 \times 10^{-2}$ | $1 \times 10^{-5} – 5 \times 10^{-5}$ | (Cadogan et al., 2014; Sriram et al., 2024) |
| **ATP_PER_H_MS** | mM ATP /(mM H$^+$·ms) | ATP cost of proton extrusion | 0.3 | 0.2–0.4 | (Lezmy et al., 2021; Sriram et al., 2024) |
| **G_K_Ca_base** | S cm$^{-2}$ | Basal Ca$^{2+}$-activated K$^+$ (KCa3.1) conductance | $1 \times 10^{-6}$ | $1 \times 10^{-7} – 5 \times 10^{-6}$ | (Cleary et al., 2024; Wang et al., 2022) |
| **K_ROS_Meff / K_ROS_Ca** | ROS per (1−Meff )/ROS units per mM Ca$^{2+}$ | Empirical ROS coupling factors | 1.0 / 2 × 10$^3$ | 0.5–2 / — | (Lezmy et al., 2021; Wang et al., 2022) |
| **D_K / D_Na / D_Cl** | μm$^2$ ms$^{-1}$ | Effective cytoplasmic diffusion | 1.1 / 1.0 / 1.2 | 0.8–1.5 | (Djurich & Secomb, 2023; Nechyporuk-Zloy et al., 2008) |
| **D_H / D_HCO3 / D_CO2** | μm$^2$ ms$^{-1}$ | Diffusion of H$^+$, HCO$_3^-$, CO$_2$ | 0.5 / 1.2 / 1.7 | 0.5–1.0 / — / — | (Endeward et al., 2017; Serowy et al., 2003) |
| **CA_GATE / PH_GATE** | M / pH | Connexin gating thresholds | $5 \times 10^{-7}$ / 7.0 | — | (Cotrina et al., 1998; Sáez et al., 2003) |
| **A_FIXED / OSM_E / TAU_WATER_MS** | mM/mOsm external/ ms | Fixed anions, extracellular osmolality, water flux τ | 155 mM / 300 mOsm / 500 ms | 150–160 / 290–310 / 400–600 | (Murakami & Kurachi, 2015; Sriram et al., 2024) |
| **ATP_OX_RATE / ATP_GLY_RATE** | mM ms$^{-1}$ | OxPhos & glycolytic ATP synthesis at extremes of Meff | 0.02 / 0.01 | — / $5 \times 10^{-6} – 1.5 \times 10^{-5}$ | (Lezmy et al., 2021; Tewari & Majumdar, 2012) |

| TAU_ROS_MS | ms | ROS decay time constant | $3 \times 10^4$ | $2-4 \times 10^4$ | (Lezmy et al., 2021; Wang et al., 2022) |
|---|---|---|---|---|---|
| G_FIELD | S cm⁻² | Electrotactic field conductance term | $5 \times 10^{-5}$ | $1 \times 10^{-6} - 1 \times 10^{-4}$ | (Dossi et al., 2024; Ma et al., 2015) |
| TAU_CLOCK_H / TAU_CC_H | h | Circadian & cell-cycle time constants | 24 / 40 | — | (Dossi et al., 2024) |
| ATP_DIV_M / VM_DIV_MV / P_MIGRATE | mM/mV /probability | Division ATP threshold, Vm threshold, migration prob. | 1.5 mM / −40 mV / 0.10 | 1.3–1.5 / −45–−35 / 0.05–0.15 | (Dossi et al., 2024; Sriram et al., 2024) |
| G_Ca_VG | S cm⁻² | Low-density VG Ca²⁺ channels | $1 \times 10^{-7}$ | $1 \times 10^{-8} - 1 \times 10^{-7}$ | (Latour et al., 2003; Zeng et al., 2009) |
| VHALF_M_Ca / K_M_Ca / TAU_M_Ca_MS | mV/mV/ ms | Activation mid-point, slope, τ | −20 mV / 6 mV / 2.5 ms | — | (Cheli et al., 2016; Scarnati et al., 2020; Van Hook et al., 2019) |
| VHALF_H_Ca / K_H_Ca / TAU_H_Ca_MS | mV/mV. ms | Inactivation parameters | −40 mV / −6 mV / 40 ms | — | (Audrey et al., 2019; Lines et al., 2023) |
| TAU_PSI_MS / PSI_REST_MV | ms / mV | Mitochondrial ΔΨm τ and resting potential | $1 \times 10^4$ / −150 | $0.8-1.2 \times 10^4$ / −160−−140 | (Gollihue & Norris, 2020; Jackson & Robinson, 2017; Rose et al., 2020) |

Appendix B. Glioblastoma Glia Cell Model Parameters

| Code constant | Unit (model) | What it is / what it does | GBM value | GBM range | Key refs |
|---|---|---|---|---|---|
| F | C mol⁻¹ | Faraday constant | 96 485.3329 | N/A | Constant physics values |
| R | J mol⁻¹ K⁻¹ | Gas constant | 8.314 462 618 | N/A | Constant physics values |

| T | K | Physiological 37 °C | 310.15 | 309–311 | Constant physics values |
|---|---|---|---|---|---|
| **RT_F** | mV | Thermal voltage (RT/F) | 26.73 | N/A | Constant physics values |
| **C_M** | µF cm$^{-2}$ | Specific membrane capacitance; reflects massive membrane ruffling in GBM | 2.95 | 1.9–4.0 | (Brandalise et al., 2020; Memmel et al., 2014) |
| **G_K_LEAK** | S cm$^{-2}$ | Inward-rectifier K$^+$ leak (Kir4.1 strongly down-regulated in GBM) | $2.55 \times 10^{-8}$ | $1 \times 10^{-9} - 5 \times 10^{-8}$ | (Brandalise et al., 2020; Takayasu et al., 2020) |
| **G_NA_LEAK** | S cm$^{-2}$ | Background Na$^+$ leak (NALCN / persistent Na$_v$) | $2.75 \times 10^{-9}$ | $5 \times 10^{-10} - 5 \times 10^{-9}$ | (eCochet-Bissuel et al., 2014; Giammello et al., 2024) |
| **G_NA_VG** | S cm$^{-2}$ | Voltage-gated Na$^+$ channels that drive depolarising spikes in GBM | $2.75 \times 10^{-8}$ | $5 \times 10^{-9} - 5 \times 10^{-8}$ | (Giammello et al., 2024; Molenaar, 2011) |
| **G_Ca_LEAK** | S cm$^{-2}$ | TRP/Orai-mediated Ca$^{2+}$ leak | $2.75 \times 10^{-9}$ | $5 \times 10^{-10} - 5 \times 10^{-9}$ | (Takayasu et al., 2020) |
| **G_Cl_LEAK** | S cm$^{-2}$ | Volume-regulated anion channel (VRAC) baseline | $2.75 \times 10^{-7}$ | $5 \times 10^{-8} - 5 \times 10^{-7}$ | (Caramia et al., 2019; McCutcheon & Spray, 2021) |
| **PUMP_MAX** | µA cm$^{-2}$ | Na$^+$/K$^+$-ATPase max current (partial pump down-regulation) | 0.65 | 0.3–1.0 | (Agrawal et al., 2023; Chen et al., 2014) |
| **I_PMCA_MAX** | µA cm$^{-2}$ | PMCA Ca$^{2+}$-pump saturation current | 0.03 | $1 \times 10^{-8} - 6 \times 10^{-7}$ | (Berrocal et al., 2025; Cong et al., 2014) |

| **G_GAP** | S cm⁻² | Connexin-43/30 gap-junction conductance between GBM cells and astrocytes | $5.5 \times 10^{-5}$ | $1 \times 10^{-5} - 1 \times 10^{-4}$ | (McCutcheon & Spray, 2021) |
|---|---|---|---|---|---|
| **ATP_BASE** | mM | Steady-state ATP pool (lowered by Warburg metabolism) | 1.0 | 0.5–1.5 | (Lee et al., 2024; Vlashi et al., 2011) |
| **ATP_KM** | mM | Michaelis constant of pumps for ATP in GBM | 0.08 | 0.04–0.12 | (Chen et al., 2014; Vlashi et al., 2011) |
| **ATP_REGEN_RATE** | mM ms⁻¹ | Basal mitochondrial + glycolytic ATP regeneration | $3.5 \times 10^{-4}$ | $2 \times 10^{-4} - 5 \times 10^{-4}$ | (Sumiyoshi et al., 2022; Vlashi et al., 2011) |
| **ATP_PER_UA_MS** | mM /(µA·ms) | ATP cost per unit ionic current (higher in GBM) | 38 | Derived | $\text{ATP\_PER\_UA\_MS} = \frac{10^3 A_{\text{cell}}}{F V_{\text{cell}}}$ |
| **AREA_CM2** | cm² | Mean surface area of a 16–18 µm GBM cell | $5.55 \times 10^{-6}$ | $3.1-8 \times 10^{-6}$ | (Brandalise et al., 2020; Memmel et al., 2014) |
| **VOL_CM3** | cm³ | Cytoplasmic volume (GBM cells are smaller) | $1.5 \times 10^{-9}$ | $1-2 \times 10^{-9}$ | (McCoy & Sontheimer, 2007; Memmel et al., 2014) |
| **K_E** | mM | Peritumoral extracellular [K⁺] (elevated) | 7.5 | 3–12 | (Benjamin et al., 2004; Strong et al., 2018) |
| **Na_E** | mM | Peritumoral [Na⁺] | 147.5 | 140–155 | (Khan et al., 2021; Wu et al., 2025) |
| **Cl_E** | mM | Peritumoral [Cl⁻] | 120 | 110–130 | (Caramia et al., 2019; McCutcheon & Spray, 2021; Trevisi & Mangiola, 2023) |
| **Ca_E** | mM | Peritumoral [Ca²⁺] | 1.35 | 1.2–1.5 | (Hartmann & Verkhratsky, 1998; Molenaar, 2011) |

| **G_NHE** | mM ms$^{-1}$ per $\Delta[H^+]$ | NHE1 proton extrusion rate (up-regulated) | $1.25 \times 10^{-4}$ | $5 \times 10^{-5} - 2 \times 10^{-4}$ | (Cong et al., 2014; McLean et al., 2000) |
|---|---|---|---|---|---|
| **KF_CO2 / KR_CO2** | see units | $CO_2$ hydration / dehydration kinetics in GBM | $2.25 \times 10^{-4}$ | $1.5–3 \times 10^{-4}$ | (Lan et al., 2017; McCoy et al., 2009) |
| **GLY_MAX** | mM ms$^{-1}$ | Max glycolytic ATP production (Warburg) | $3 \times 10^{-6}$ | $1–5 \times 10^{-6}$ | (Sumiyoshi et al., 2022) |
| **ATP_PER_H_MS** | mM ATP/(mM H$^+$·ms) | ATP cost of proton extrusion | 0.65 | 0.3–1.0 | (Agrawal et al., 2023; Cong et al., 2014) |
| **G_K_Ca_base** | S cm$^{-2}$ | Basal KCa conductance (BK channel up-regulation) | $2.55 \times 10^{-6}$ | $1 \times 10^{-7} - 5 \times 10^{-6}$ | (Brandalise et al., 2020; Wawrzkiewicz-Jałowiecka et al., 2020) |
| **K_ROS_Meff / K_ROS_Ca** | ROS per (1−Meff)/ ROS units per mM Ca$^{2+}$ | Empirical ROS-Ca$^{2+}$ coupling constants | $1.25 / 3 \times 10^2$ | $0.5–2 / 2–4 \times 10^2$ | (Agrawal et al., 2023; Sumiyoshi et al., 2022) |
| **D_K / D_Na / D_Cl** | µm² ms$^{-1}$ | Effective diffusion in crowded GBM cytoplasm | 0.45 / 0.55 / 0.55 | 0.3–0.7 | (Server et al., 2009; Shayganfar et al., 2020) |
| **D_H / D_HCO3 / D_CO2** | µm² ms$^{-1}$ | H$^+$, HCO$_3^-$, CO$_2$ diffusion | 0.115 / 0.6 / 1.5 | 0.08–0.15 / 0.4–0.8 / 1.0–2.0 | (McCoy et al., 2009; Syková & Nicholson, 2008) |
| **CA_GATE / PH_GATE** | M / pH | Connexin gating thresholds in acidic GBM milieu | $5 \times 10^{-7}$ / 6.35 | $4–6 \times 10^{-7}$ / 6.2–6.5 | (McCutcheon & Spray, 2021; Paunikar & Tamagnone, 2024) |

| **A_FIXED / OSM_E / TAU_WATER_MS** | mM/mOsm external/ms | Fixed anions, extracellular osmolality, water-flux $\tau$ | 150 mM / 310 mOsm / 450 ms | 140–160 / 290–330 / 100–800 | (Solenov et al., 2004; Varricchio & Yool, 2023) |
|---|---|---|---|---|---|
| **ATP_OX_RATE / ATP_GLY_RATE** | mM ms$^{-1}$ | OxPhos & glycolytic ATP synthesis extremes | $2.5 \times 10^{-5}$ / $1 \times 10^{-5}$ | $1$–$4 \times 10^{-5}$ / $5 \times 10^{-6}$–$1.5 \times 10^{-5}$ | (Sumiyoshi et al., 2022; Vlashi et al., 2011) |
| **TAU_ROS_MS** | ms | ROS decay $\tau$ (slower in GBM) | $5 \times 10^{4}$ | $1 \times 10^{4}$–$9 \times 10^{4}$ | (Agrawal et al., 2023; Cifarelli et al., 2021) |
| **G_FIELD** | S cm$^{-2}$ | Empirical electrotactic conductance term | $5.05 \times 10^{-5}$ | $1 \times 10^{-6}$ – $1 \times 10^{-4}$ | (Sumiyoshi et al., 2022; Zanoni et al., 2022) |
| **TAU_CLOCK_H / TAU_CC_H** | h | Circadian & cell-cycle time constants | 24 / 40 | 22–26 / 20–60 | (Sumiyoshi et al., 2022) |
| **ATP_DIV_M / VM_DIV_MV / P_MIGRATE** | mM/mV/probability | ATP threshold for division, Vm threshold, migration prob. | 1.4 mM / −40 mV / 0.20 | 1.2–1.6 / −50–−30 / 0.1–0.3 | (Chen et al., 2014; Zanoni et al., 2022) |
| **G_Ca_VG** | S cm$^{-2}$ | Sparse VG Ca$^{2+}$ channel density | $5.5 \times 10^{-8}$ | $1 \times 10^{-8}$–$1 \times 10^{-7}$ | (Giammello et al., 2024; Molenaar, 2011) |
| **VHALF_M_Ca / K_M_Ca / TAU_M_Ca_MS** | mV/mV/ms | Activation midpoint, slope, $\tau$ | −35 mV / 7 mV / 2.5 ms | -40.0--30.0 / 6.0–8.0 / 1.0–4.0 | (Giammello et al., 2024; Molenaar, 2011) |
| **VHALF_H_Ca / K_H_Ca / TAU_H_Ca_MS** | mV/mV/ms | Inactivation parameters | −55 mV / −7 mV / 32.5 ms | -65.0–-45.0 / -8.0–-6.0 / | (Coronas et al., 2020; Dube et al., 2024; Zhang et al., 2017) |

| | | | | 15.0–50.0 | |
|---|---|---|---|---|---|
| **TAU_PSI_MS / PSI_REST_MV** | ms / mV | Mitochondrial $\Delta\Psi$m $\tau$ and resting potential (hyperpolarised) | $3.25 \times 10^4$ / $-175$ | $5 \times 10^3$–$6 \times 10^4$ / $-220$ – $-130$ | (Sumiyoshi et al., 2022; Zhang et al., 2023) |